\documentclass[12pt,a4paper]{article}
\pdfoutput=1

\usepackage{graphicx,amssymb,amsmath,color}
\usepackage[linkcolor={blue},citecolor={red},colorlinks=true]{hyperref}
\usepackage{enumerate}
\usepackage[margin=2cm]{geometry}
\usepackage{slashed}
\usepackage{amsmath} 
\usepackage{amsfonts}
\usepackage{amssymb}
\usepackage{graphicx, rotating}
\usepackage{epstopdf}
\usepackage{epsfig}
\usepackage{braket}
\usepackage{empheq}
\usepackage{latexsym}
\usepackage{graphicx}
\usepackage{subcaption}
\usepackage{color}
\usepackage{amsmath,amssymb}
\usepackage{cite}
\usepackage{slashed}
\usepackage{hyperref}
\usepackage{xcolor}
\usepackage{multirow}

\newcommand{\be}{\begin{equation}}
\newcommand{\ee}{\end{equation}}

\def\beq{\begin{equation}}
\def\eeq{\end{equation}}
\def\bea{\begin{eqnarray}}
\def\eea{\end{eqnarray}}

\def\bit{\begin{itemize}}
\def\eit{\end{itemize}}

\def\l{\left}
\def\r{\right}

\def\baa{\begin{array}}
\def\eaa{\end{array}}

\def\d{\partial}

\def\simgt{\mathrel{\lower2.5pt\vbox{\lineskip=0pt\baselineskip=0pt
           \hbox{$>$}\hbox{$\sim$}}}}
\def\simlt{\mathrel{\lower2.5pt\vbox{\lineskip=0pt\baselineskip=0pt
           \hbox{$<$}\hbox{$\sim$}}}}

\def\bfc{\begin{figure}\begin{center}}
\def\efc{\end{center}\end{figure}}
\def\nn{\nonumber\\}

\hypersetup{colorlinks, citecolor=bluscuro, linkcolor=black, urlcolor=bluscuro}
\definecolor{rossos}{cmyk}{0,1,1,0.55}
\definecolor{bluscuro}{rgb}{0.15, 0.2, .85}
\definecolor{bluchiaro}{cmyk}{1,.3,0.,0.1}

\usepackage{multicol}

\newcommand{\bc}{\begin{center}}
\newcommand{\ec}{\end{center}}

\newcommand{\blu}{\color{blue}}

\newcommand{\bag}{\begin{align}}
\newcommand{\eag}{\end{align}}

\usepackage{soul}

\begin{document}
\begin{flushright}
\hspace{3cm} 
SISSA 28/2019/FISI
\end{flushright}
\vspace{.6cm}
\begin{center}

\hspace{-0.4cm}{\Large \bf 
Gravitational traces of broken gauge symmetries}\\[0.5cm]

\vspace{1cm}{Aleksandr Azatov$^{a,b,c,1}$, Daniele Barducci$^{d,e,2}$ and Francesco Sgarlata$^{a,b,c,3}$}
\\[7mm]
 {\it \small

$^a$ SISSA International School for Advanced Studies, Via Bonomea 265, 34136, Trieste, Italy\\[0.15cm]
$^b$ INFN - Sezione di Trieste, Via Bonomea 265, 34136, Trieste, Italy\\[0.1cm]
$^c$ IFPU, Institute for Fundamental Physics of the Universe, Via Beirut 2, 34014 Trieste, Italy\\[0.1cm]

$^d$ Universit\`a degli Studi di Roma la Sapienza, Piazzale Aldo Moro 5, 00185, Roma, Italy\\[0.15cm]
$^e$ INFN - Sezione di Roma, Piazzale Aldo Moro 5, 00185, Roma, Italy\\[0.1cm]
 }

\end{center}

\bigskip \bigskip \bigskip

%%%%%%%%%%%%%%%%%%%%%%%%%%%%%%%%%%%%%%%%%%%%%%%%%%%%%%%%%%%%%%%%%%%%%%%%%%
\centerline{\bf Abstract} 
\begin{quote}
We investigate first-order phase transitions arising from hidden sectors, which are in thermal equilibrium with the Standard Model bath in the Early Universe. Focusing on two simplified scenarios, a higgsed $U(1)$ and a two scalar singlet model, we show the impact of friction effects acting on the bubble walls on the gravitational wave spectra and the consequences for present and future interferometer experiments. We further comment on the possibility of disentangling the properties of the underlying theory featuring the first-order phase transition should a stochastic gravitational-wave signal be discovered. 
\end{quote}

\vfill
\noindent\line(1,0){188}
{\scriptsize{ \\ E-mail:
\texttt{$^1$\href{mailto:aleksandr.azatov@NOSPAMsissa.it}{aleksandr.azatov@sissa.it}},
\texttt{$^2$\href{mailto:daniele.barducci@NOSPAMroma1.infn.it}{daniele.barducci@roma1.infn.it}}, 
\texttt{$^3$\href{mailto:fsgarlat@NOSPAMsissa.it}{francesco.sgarlata@sissa.it}}
}}

\newpage
\tableofcontents

\section{Introduction}

The recent detection of gravitational waves (GW) signals originating from the merging of black holes~\cite{Abbott:2016blz} and neutron stars binaries\footnote{A candidate event arising from the merging of a black hole and a neutron star have recently been observed as reported in the Gravitational Wave Candidate Event Database, see~\url{https://gracedb.ligo.org/superevents/S190814bv/}.}\cite{GBM:2017lvd} have provided a remarkable test for Einstein's theory of gravity, so far in excellent agreement with experimental data.

Remarkably, the enormous progress that has been made in  GW observational cosmology also provides a new array of experimental tests in the search for new physics beyond the Standard Model (SM). The prince example is one of models featuring a first-order phase transition (FOPT). In the SM, neither the QCD \cite{Bernard:2004je,Aoki:2006we,Cheng:2006qk} nor the electroweak (EW) phase transitions are first-order\cite{Kajantie:1996mn,Rummukainen:1998as,Csikor:1998eu}. Yet, the presence of a FOPT is one of the possibilities to achieve a departure from thermal equilibrium during the cosmological evolution of the Universe, thus satisfying one of the three Sakharov conditions~\cite{Sakharov:1967dj} necessary to create a baryon asymmetry in the present Universe. It is well known that a FOPT leads to a  stochastic GW background signal~\cite{Witten:1984rs}.  Interestingly, current and future GW interferometer experiments are (and will be) sensitive to a vast range for the scale of the FT, ranging from $\sim1\;$GeV to $\mathcal{O}(10^{10})\;$GeV, thus making this direction of experimental searches complementary to others in the experimental particle physics program. 

These experimental advances in turn call for analogous progresses 
in the theory community. 
It is necessary to have precise calculations of the GW spectrum
sourced by a FOPT in order to fully exploit the potential of the various ongoing and planned
experiments. 
On the general ground, the stochastic GW signal arising from a FOPT gets contributions from three different sources: a) collision of true 
vacuum bubbles expanding in the false vacuum background 
\cite{Cutting:2018tjt,Kosowsky:1991ua, Kosowsky:1992vn,Kosowsky:1992rz,
Jinno:2017fby} b) sound waves of the plasma \cite{Hindmarsh:2017gnf,Konstandin:2017sat,Jinno:2017fby} and c) turbulent motion of the plasma itself \cite{Caprini:2015zlo,Kamionkowski:1993fg,Kosowsky:2001xp,Caprini:2009yp}.
Accurate theory predictions are necessary in order 
to explore the inverse problem, {\emph{i.e.}} the extraction of the 
properties of the underlying new physics theory should a GW signal associated with a FOPT be discovered.

Of course, the precise solution to the problem is still far from being achieved. Currently, the shape and the amplitude of the GW signals are known only in some limited ranges of the parameters controlling a FOPT, and various effect can modify the predictions for the GW spectrum. 
In the case of relativistic moving bubble walls, an important factor that affects the relative importance of the various contributions to the stochastic GW background is whether they reach a terminal velocity or not before they collide. Expanding bubbles experience, in fact, a friction force due to the surrounding plasma and early studies showed that these forces were independent of the Lorentz factor $\gamma$ of the bubble wall. Thus, whenever the driving force is exceeding the inward pressure, relativistic bubbles will keep accelerating until they collide~\cite{Bodeker:2009qy}. 
However, a more recent calculation \cite{Bodeker:2017cim} took into account higher-order friction effects on the bubble wall, hereafter next to leading order (NLO) friction. The authors of \cite{Bodeker:2017cim} showed indeed that in the presence of particles with phase-dependent masses, the friction force is proportional to the $\gamma$ factor itself, thus causing the bubble wall to approach a terminal velocity.  If this terminal velocity is reached before the bubbles collision, the contribution to the GW spectrum arising from this source turns out to be almost completely irrelevant. In this case, the signal is dominated by sound waves and turbulence contributions. Since these sources have different peak frequencies and power-law scalings, the resulting total spectrum turns out to be drastically different. In the presentation of our numerical results, we will, however, neglect the turbulence contribution, as recommended by the latest update from the LISA cosmology working group~\cite{Caprini:2019egz}, due to the uncertainties associated with the calculation of this effects. We will nevertheless qualitatively comment on how our results are modified when this contribution is added to the total GW spectrum.
The results of this effect is anticipated in Fig.~\ref{fig:intro}, where we show the expected total power spectrum from the FOPT with and without the inclusion of NLO friction effect overlaid with the expected sensitivity of various GW experiments.
 The two curves  are obtained by tuning all of the parameters controlling the FOPT (except the amount of NLO friction effect) to be the same (the definitions of these parameters can be found in Eqs.~(\ref{eq:rad}), (\ref{eq:alpha}) and  (\ref{eq:Treh})). Interestingly, we can observe that different models with similar potentials at the time of the FOPT can 
have a very distinct GW signal if only one of the two is subject to the NLO friction effect 
above described.

\bfc
\includegraphics[scale=0.7]{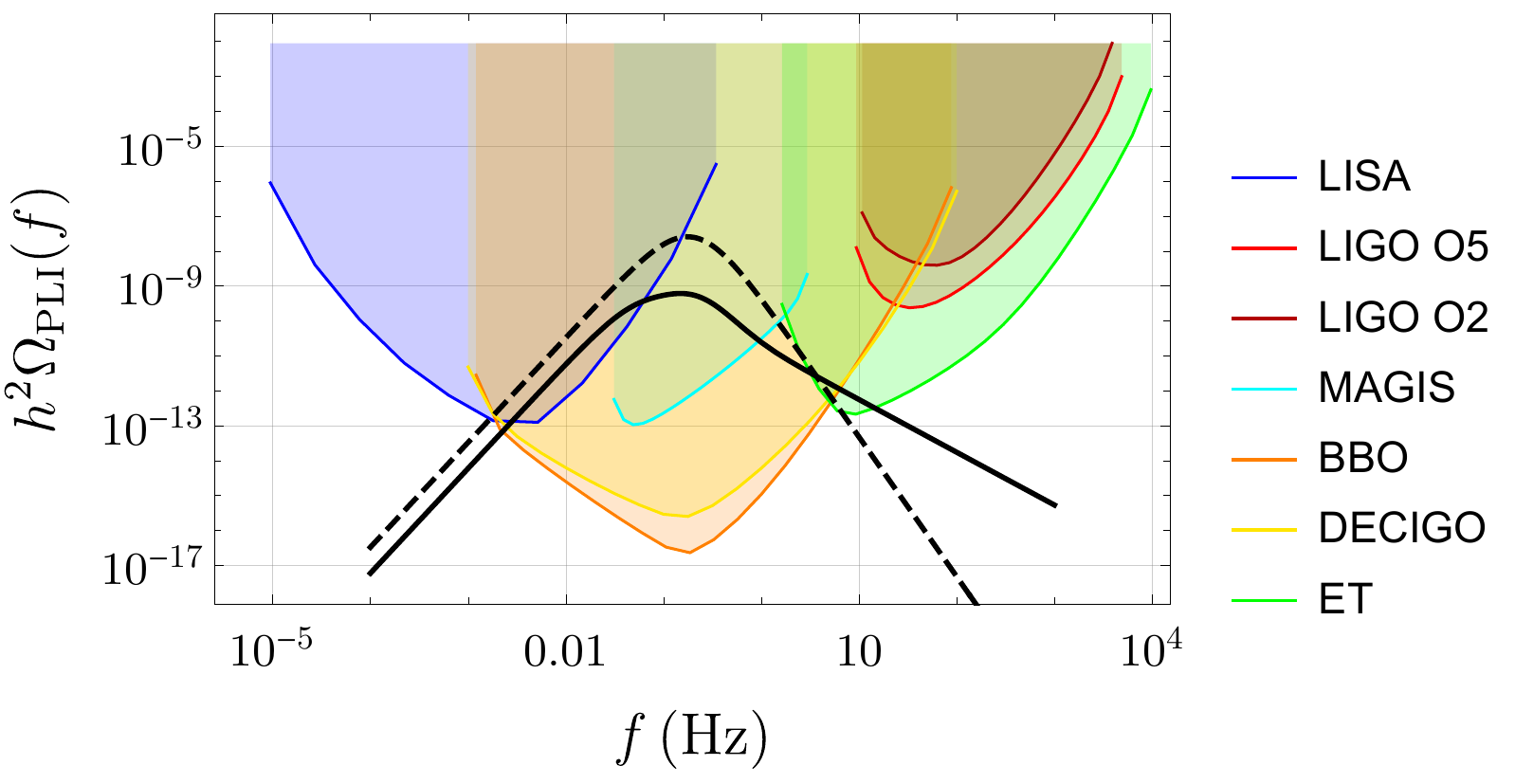}
\caption{{\it Typical GW signal expected from a FOPT without (solid) and with (dashed)
the inclusion of the NLO effect of ~\cite{Bodeker:2017cim} reviewed in Sec.~\ref{sec:dynamics}. We fix $\alpha=10,\alpha_{\infty}=1, H_{reh} R_*=10, T_{reh}=10^5$ GeV. }
\label{fig:intro}}
\efc

In this paper, we discuss two simplified scenarios that can present a FOPT leading to a stochastic GW signal: a classical scale-invariant model with a higgsed 
$U(1)$ symmetry and a model with two singlet scalar fields. We compute in detail the FOPT parameters affecting the GW spectrum for the two cases and show how the presence of the NLO friction force affects the reach of current and future experiments. 
We further show that in certain regions of the parameter spaces, the two models can exhibit a similar effective potential, producing however different features in the GW 
spectrum due to the different impact of the NLO friction force. This allows for a partial determination of the underlying theory particle content in case of a stochastic GW background signal discovery~\footnote{The feasibility of discriminating an underlying model or part of its properties from the properties of the GW spectra is subjects of various articles, see {\emph{e.g.}}~\cite{Croon:2018erz,Alanne:2019bsm}.
}. 
 
Before our analysis, there were various studies looking at similar simplified scenarios featuring a FOPT as the ones subject of our study. For example, ~\cite{Breitbach:2018ddu} analyzed the same simplified models, focusing 
however to regions of the parameter space where the NLO friction effect 
of~\cite{Bodeker:2017cim} was indeed negligible, while~\cite{Bian:2019szo} and~\cite{Hashino:2018zsi} analyzed a classically conformal $U(1)_{B-L}$ model and a higgsed $U(1)$ extension of the SM respectively, without discussing the friction effects. Ref.~\cite{Fairbairn:2019xog} included instead the friction effects. That work analyzed a model similar to the one we study in Sec.~\ref{sec:model_1}, based however on an underlying $SU(2)$ gauge symmetry~\footnote{GW signals from the FOPT within a similar class of models have also been recently considered in \cite{Mohamadnejad:2019vzg,Dev:2019njv}. }.
 All together, our work is complementary to existing analyses since we aim at investigating toy models with very similar potential at the moment of the FOPT leading however, to different experimental signatures. While we refrain from performing a statistical analysis aimed at determining to which extent two theories can be disentangled, our study aims at pointing out that the spectral shape of a detected signal can, in principle, be used to infer some properties of the underlying theory featuring a FOPT. 

The paper is organized as follows. In Sec.~\ref{sec:model_1} we introduce the higgsed $U(1)$ model and describe the general conditions for a successful phase transition. In Sec.~\ref{sec:dynamics} we review the dynamics of bubble expansions, discuss the friction effects that can affect the GW spectrum, and illustrate the present and future experimental coverage on the higgsed $U(1)$ model. Then in Sec.~\ref{sec:model_2} we discuss the comparison between the higgsed $U(1)$ model and the one with two scalar fields. We then conclude in Sec.~\ref{sec:concl}.

\section{The simplest toy model}
\label{sec:model_1}
We now focus on a simple perturbative toy model with a massless complex scalar field and a  gauged $U(1)$ symmetry which exhibits a FOPT induced by quantum corrections. This is a classically scale-invariant theory described by
\begin{equation}
\label{eq:lag-model_1}
	\mathcal{L} = -\frac{1}{4}F_{\mu\nu}^2 + D_\mu \phi^\dagger D^\mu \phi - V_1(\phi,T)
	\end{equation}
where $D^\mu\phi = \partial^\mu\phi -i g A^\mu \phi$ and the potential $V_1(\phi,T)$ is generated at one-loop level (for simplicity we have set the tree level potential to be equal to zero). In the real scalar fields basis we choose $\phi = \frac{1}{\sqrt{2}}\left(\phi_1+i\phi_2\right)$, where $\phi_1$ parametrises the direction along which the field takes its vacuum expectation value (VEV).

At zero temperature, the one-loop contribution is the Coleman-Weinberg effective potential~\cite{Weinberg:1973am}
\begin{equation}
\label{CWpotential}
	V_{\rm{CW}} = \sum_i g_i \frac{m_i^4}{64\pi^2}\left[\log\left(\frac{m_i^2}{\mu_R^2}\right) -c_i\right]
\end{equation}
where the sum runs over the different degrees of freedom of the model. In the case at hand, the latter are the longitudinal and transverse components of $A_\mu$ as well as two scalar degrees of freedom from $\phi$. The coefficients $g_i$ count the degrees of freedom for each species and $c_i = 3/2,5/2$ for scalars and vectors respectively. The scale $\mu_R$ is a renormalization scale, in terms of which the physical parameters of the model are matched and we are free to choose the parametrization $\mu_R = g w$, where $w$ has dimension of a VEV. 
In this model, where the classical potential is tuned to zero, at $T=0$ only a field-dependent mass for the gauge field is generated
\begin{equation}
	m_A^2(\phi_1) = g^2\phi_1^2\,,
\end{equation}
and $V_1(\phi,T)$ develops two symmetric global minima at $|\phi_1| \sim 2.7\, w$ while the point $\phi_1=0$ becomes a local maximum. 

When the system interacts with a thermal bath, thermal corrections to the one-loop effective potential become important and at high temperatures the local maximum at $\phi_1=0$ may turn into a local minimum (false vacuum). A potential barrier is thus generated between $\phi_1=0$ and the global minimum $\phi_1\sim 2.7 w$. This leads to the possibility of a FOPT.  Thermal corrections can be taken into account by adding the  following terms to the Coleman-Weinberg effective potential
\begin{equation}
	V_T(\phi) = \sum_i \frac{g_i}{2\pi^2}T^4 J\left(\frac{m_i^2(\phi)}{T^2}\right)\,,\qquad J(y^2) = \int_0^{+\infty}dx \, x^2 \log\left[1 - \text{exp}\left(-\sqrt{x^2+y^2}\right)\right]
\end{equation}
where $J(y^2)$ enjoys the asymptotic expansions ~\cite{Curtin:2016urg}
\begin{align}
\label{eq:jbexpansion}
	J(y^2\ll 1) = -\frac{\pi^4}{45}+\frac{\pi^2}{12}y^2-\frac{\pi}{6}y^3+\cdots\,,\qquad J(y^2\gg 1) = -\sum_{n=1}^{m\geqslant 3}\frac{1}{n^2}y^2 K_2(y\cdot n)
\end{align}
and in the last equation we have introduced the second-kind Bessel functions $K_2(z)$.
On top of this correction, in order to have a reliable prediction for the total potential,  it is necessary to include higher order effects described by the contributions of the so called daisy diagrams. This can be easily done using the Truncated Full Dressing procedure~\cite{Curtin:2016urg}, where the one-loop effective potential is modified to be
\bea
\label{eq:pottotal}
&&	V_{eff}(\phi,T) = 
	V_{CW}\left(m_i^2 + \Pi_i^2\right) + V_T\left(m_i^2+ \Pi_i^2\right) \nn
	&&\Pi^2_{\phi_{1,2}} =\frac{ g^2 T^2}{4}\,,\qquad\Pi_{\text{Transv}(A)}^2=0\,,\qquad \Pi_{\text{Long}(A)}^2 = \frac{g^2T^2}{3} \ .
\eea

We  note that the size of the $n+1$ over $n$ thermal loop corrections to the potential scales as the coupling $\sim g$~\cite{Weinberg:1974hy,Arnold:1992rz}. For this reason we will restrict our analysis only to the region where $g\lesssim 1$.

In order to complete the discussion of the thermal correction to the potential we need to know  the relation between the temperature of the dark sector described by Eq.~\eqref{eq:lag-model_1} and that of the SM. 
 Throughout our analysis we will implicitly assume that the two sectors are in thermal equilibrium. This can be easily achieved by assuming, {\emph{e.g.}}, an Higgs portal like interaction between the two sectors 
 \be
 \label{eq:portal}
 \lambda_{{\rm mix.}} |H^2||\phi|^2. 
 \ee
Interactions of these type are subject to various kind of constraints and have been widely studied in the literature. The main bounds on this portal like interaction  come from the existence of stable relics in the dark sector and the presence of additional relativistic degrees of freedom in the Early Universe~\cite{Fairbairn:2019xog,Breitbach:2018ddu}. These bound can however be easily avoided by introducing a kinetic mixing between the hypercharge $U(1)_Y$ and the dark $U(1)$ groups and by assuming that the hidden sector lies at a scale which is at least one order of magnitude higher than the EW one, which we will always assume to be the case. On the other side if the hidden sector is light, the mixing parameter should be necessarily small, which means that the hidden sector and SM might not be in thermal equilibrium leading to interesting signals. We refer to~\cite{Fairbairn:2019xog,Breitbach:2018ddu} for a more detailed study of the interaction of Eq.~\eqref{eq:portal} and assume throughout the paper that the SM and the dark sector are in thermal equilibrium and that relevant constraints can be easily satisfied.

\subsection{Transition rates and percolation}
\label{sec:percolation}
\begin{figure}[t]
\begin{center}
\includegraphics[height=5.5cm]{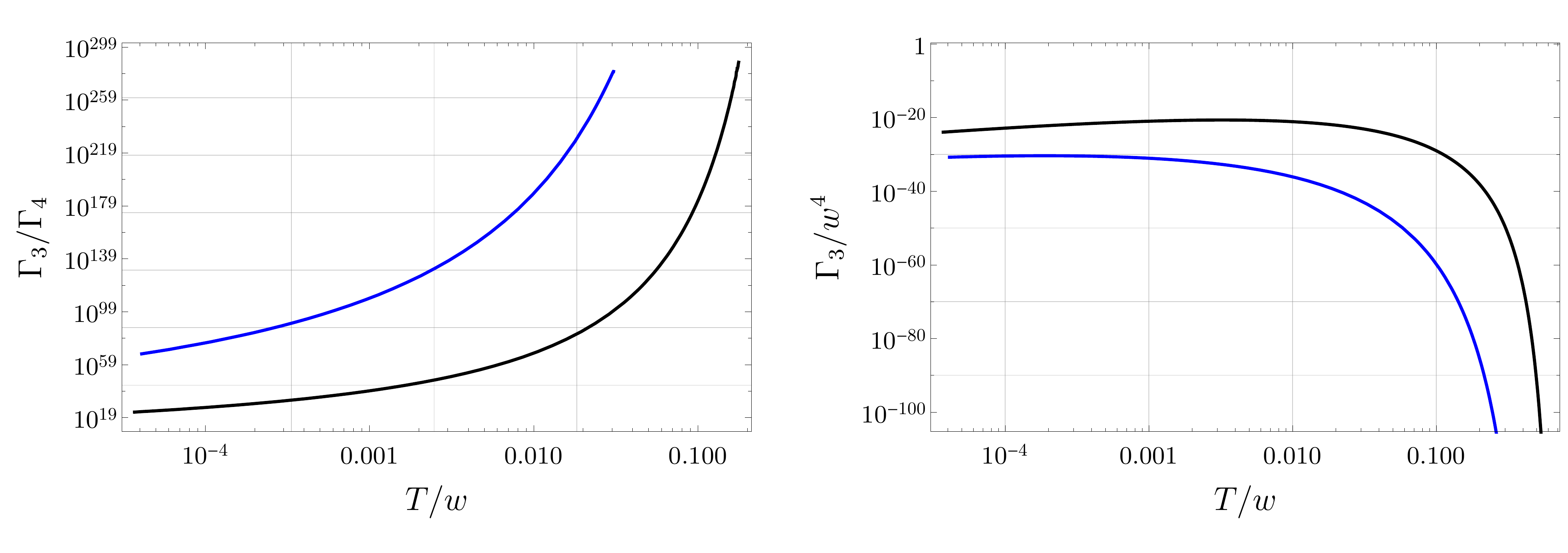}
\caption{
{\it Left: Ratio between the  $O(3)$ and $O(4)$ decays rates, as defined in Eq.~\eqref{eq:rates}.  The $O(3)$ decay rate generically dominates for non-vanishing temperatures. Right:  $O(3)$ decay rate as function of the temperature. In both panels we plot the values for $g=0.8$ (blue) and $g=1$ (black).}}
\label{fig:decayrates}
\end{center}
\end{figure}

In field theory, phase transitions are driven by bubbles nucleation\cite{Coleman:1977py} and subsequent percolation around all the available volume. In the Euclidean space, the bubbles correspond to stationary minimum-energy solutions interpolating the false and true vacuum. The decay rate into the true vacuum gets leading contributions from $O(3)$ and $O(4)$ symmetric bounce  solutions \cite{Coleman:1977py,Linde:1980tt,Linde:1981zj}
\begin{equation}
\label{eq:rates}
\Gamma(T) \simeq \Gamma_3(T) + \Gamma_4(T)= T^4 \left(\frac{S_3}{2\pi T}\right)^{3/2}e^{-S_3/T}+ \frac{1}{R_0^4}\left(\frac{S_4}{2\pi}\right)^2 e^{-S_4} ,
\end{equation}
where $S_{n}$ is the Euclidean actions for the $O(n)$ case, $R_0$ is the bubble 
radius and the first addend corresponds to thermal transition whereas the latter to 
quantum tunnelling. For our study we have calculated the
 bounce solutions numerically implementing an overshoot/undershoot method in {\tt Mathematica}, eventually crosschecking our results with the ones obtained with the 
 {\tt Cosmotransition} package~\cite{Wainwright:2011kj}.
  We have found that for the model under 
consideration the phase transition is always dominated by  thermal fluctuations and 
that quantum tunneling effects are largely negligible, as shown in the left panel of 
Fig.~\ref{fig:decayrates}. In the right panel  we show instead how the vacuum decay 
rate evolves with the temperature during the Early Universe evolution due to the 
modification of the temperature dependent part of the effective potential.

Various temperatures are defined in order to characterize the phase transition~\cite{Ellis:2018mja,Caprini:2015zlo,Espinosa:2010hh}. We start with 
the  nucleation temperature $T_n$, defined  by the condition of  one bubble  nucleation per Hubble volume 
\begin{equation}
\label{nuclCond}
1 = \int_{t_c}^{t_n} dt \frac{\Gamma(t)}{H(t)^3} = \int_{T_n}^{T_c} \frac{dT}{T}\frac{\Gamma(T)}{H(T)^4}, 
\end{equation}
where $H(T)$ is the Hubble constant and $T_c$ is the critical temperature, {\emph{i.e.}} the temperature at which the two minima are degenerate. For this toy model, the critical temperature scales roughly with the gauge coupling $T_c/w \sim 0.85 g $. The Hubble constant instead takes the usual form
\begin{equation}
\label{HubbleConstant}
H(T)^2 = \frac{1}{3M_{pl^2}}\left(\rho_R + \Delta V(T)\cdot \mathcal{P}_{false}(T)\right)\,,\qquad \rho_R=\frac{\pi^2g_*}{30}T^4
\end{equation}
where $\Delta V(T)$ is the potential difference between the false and the true vacuum and $\mathcal{P}_{false}(T)$ is the ratio of volume trapped in the false vacuum. Note that the difference of potential {\emph{i.e.}} the latent heat will be related to the effective potential defined in Eq.~\eqref{eq:pottotal} as 
\be\label{eq:deltaVdv}
\Delta V(T)=\Delta V_{eff} -\frac{T}{4}\frac{ \d\Delta V_{eff}}{\d T} \ .
\ee
However, numerically the part containing the derivative turns out to be completely irrelevant numerically.
When the potential barrier persists at $T=0$, the vacuum energy density dominates the Hubble constants and drives an exponential expansion, therefore its contribution to the nucleation condition Eq.~(\ref{nuclCond}) becomes relevant and cannot be neglected.
Since the integral in Eq.~(\ref{nuclCond}) gets dominant contributions only around $T=T_n$ we can estimate the nucleation condition as
\begin{equation}
\Gamma(T_n) \sim H(T_n)^4
\end{equation}
which serves as definition for the nucleation temperature. For fast PTs, the temperature at which bubbles percolate and fill the Universe (\emph{i.e.} the percolation temperature $T_p$) is commonly taken equal to the nucleation temperature $T_n\sim T_p$. However, for PTs with strong supercooling ({\emph{i.e.}} slow phase transitions), the vacuum energy drives an exponential expansion and therefore the transition could not be efficient, as nucleated bubbles may never meet \cite{Ellis:2018mja}. In this case, efficient transitions will be completed at temperatures substantially lower than the nucleation temperature and a more precise condition can be imposed by requiring that the space volume of points still trapped in the false vacuum definitely decreases with time. Recently all of this issues have been carefully discussed in~\cite{Ellis:2018mja} and here for the clarity we briefly review all of the necessary conditions.
 The probability to find a point in the false vacuum (which is exactly the ratio between the false and true vacuum appearing in the Eq. \eqref{HubbleConstant}) is given by  \cite{Guth:1979bh,Guth:1981uk}:
\begin{equation}
\label{IFunction}
\mathcal{P}_{false}(T)= e^{-I(T)}, ~~~I(t) = \frac{4\pi}{3}\int_{t_c}^t dt' \,\Gamma(t') a(t')^3 r(t,t')^3
\end{equation}
where $a(t)$ is the Friedmann-Robertson-Walker scale factor and $r(t,t') = 
\int_t^{t'}d \tilde{t}\,v_w/a(\tilde{t}) d\tilde t$ is the comoving radius at the time 
$t$ of a bubble nucleated at time $t'$ propagating with speed $v_w$. Assuming 
adiabatic expansion, we can convert Eq.~(\ref{IFunction}) into an integral over 
temperatures by means of the adiabatic time-temperature relation $dt/dT = -\left(TH(T)\right)^{-1}$; the final expression is \cite{Ellis:2018mja}
\begin{equation}
\label{eq:percolation}
I(T) = \frac{4\pi}{3}\int_T^{T_c} \frac{dT'\, \Gamma(T')}{H_V T'^4 \chi(T')}\left(\int_T^{T'} \frac{d \tilde{T}}{H_V \chi(\tilde{T})}\right)^3
\end{equation}
where $\chi(T) = \sqrt{1+\rho_R/\Delta V(T)}$ and $H_V^2 = \Delta V/(3M_{pl}^2)$.
\begin{figure}[t]
\begin{center}
\includegraphics[height=6cm]{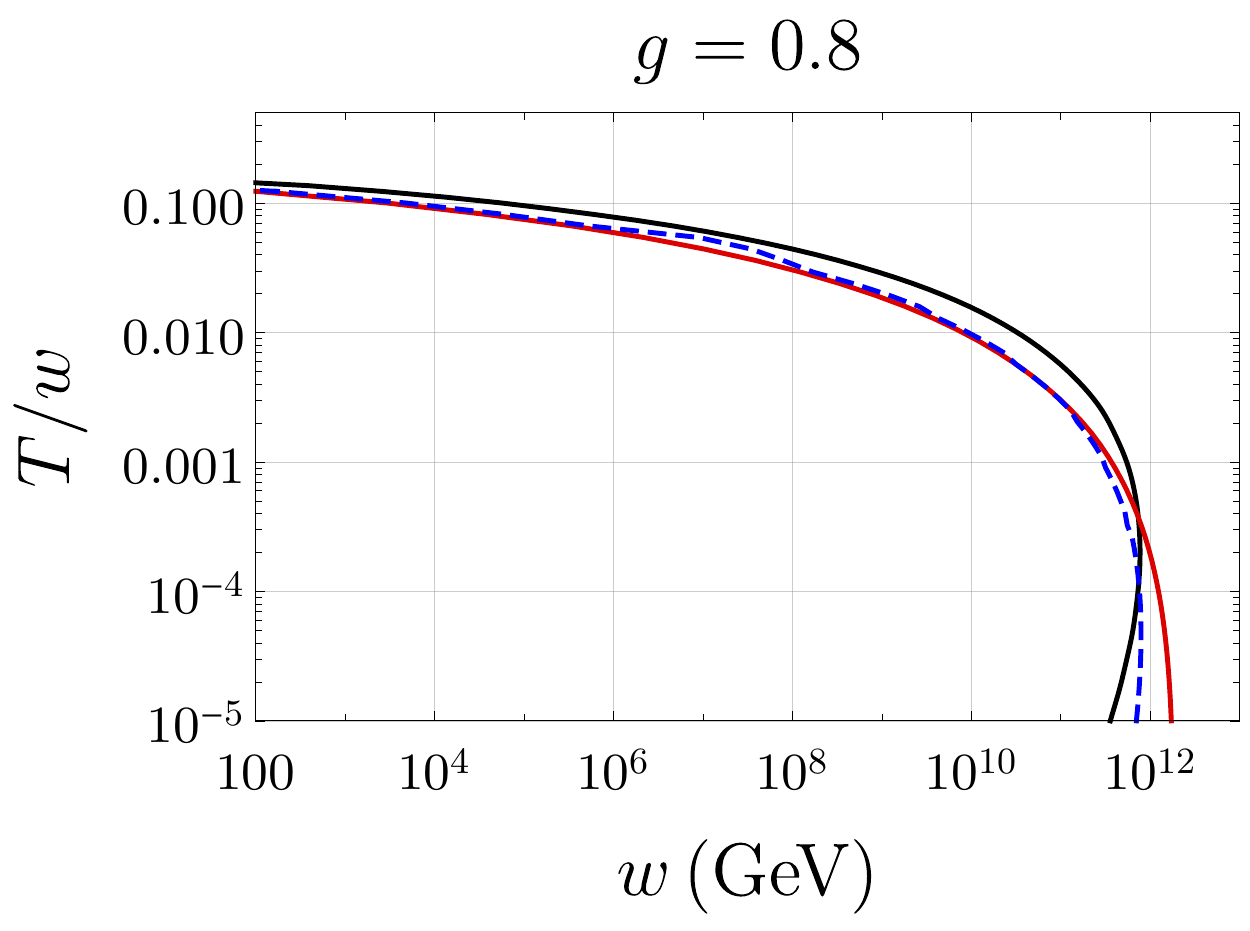}\,\,\,\,\,\,
\includegraphics[height=6cm]{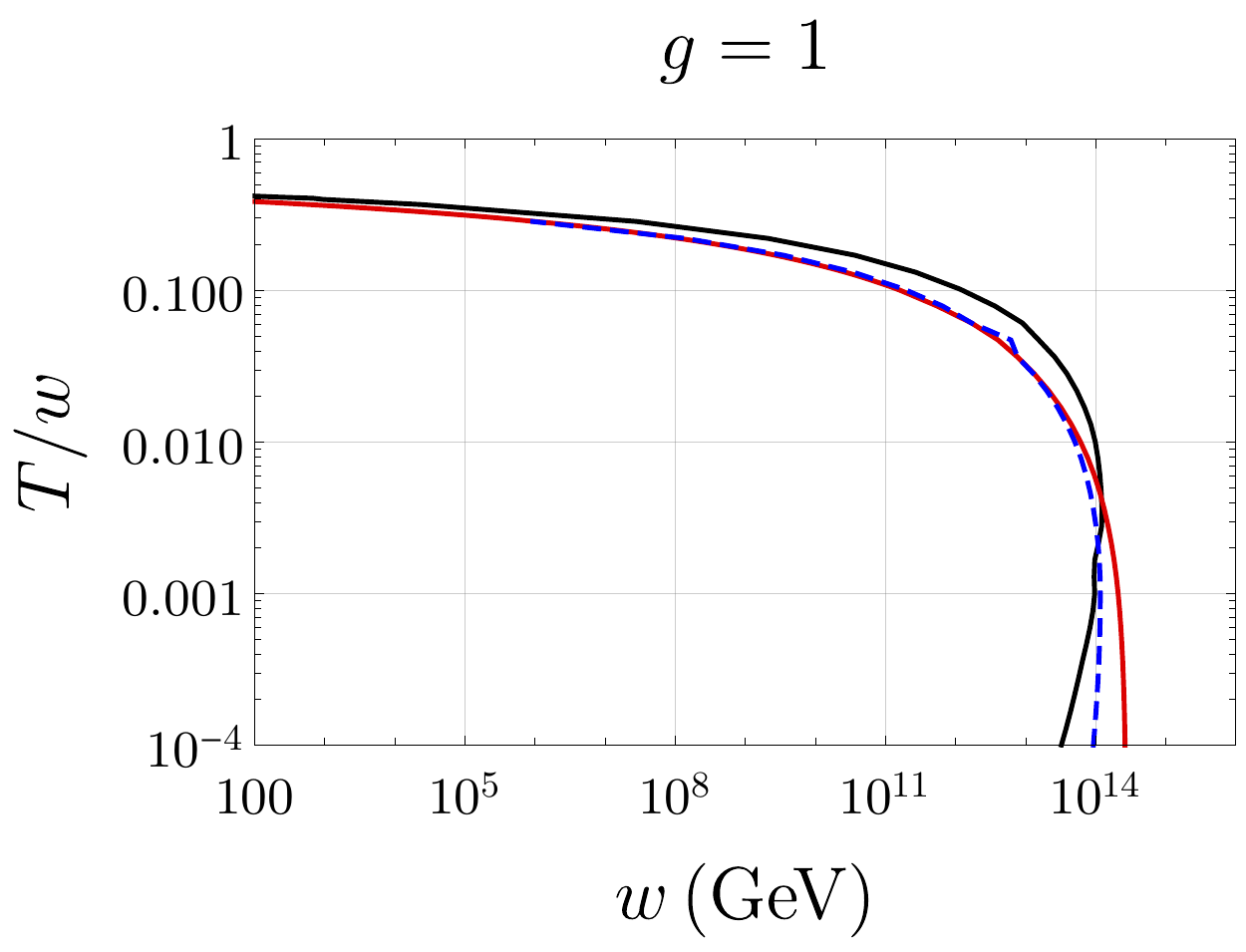}
\caption{{\it Nucleation (black) and percolation conditions for relativistic bubbles with $v_w=1$. The red line corresponds to $I(T_p/w)=0.34\,w^4$ while the blue dashed line to the more accurate condition of decreasing false vacuum volume in Eq.~(\ref{strongPerCond}). We anticipate here that non-relativistic bubbles expansion gives relevant phenomenological contributions for small values of $w$, see Fig.~\ref{fig:boundsmodel1}. In this region however the function $I(T_p/w)$ weakly depends on $v_w$ and thus, for non-relativistic bubbles, the percolation temperature does not significantly change.
}}
\label{fig:percolationConditions}
\end{center}
\end{figure}
 In radiation-domination scenario, it is considered to have a successful percolation when $I(T_p)\geq 0.34$~\cite{Guth:1982pn}. This condition serves then as definition of the percolation temperature. However,
in inflationary scenario,  it is possibile that the false vacuum volume inflates and bubbles never meet. For example, assuming a constant $\Gamma(T)$ and vacuum-dominated expansion, the expression Eq.~(\ref{IFunction}) evaluated at infinite time $t\gg t_c$ becomes
\begin{equation}
I(t) = \frac{4\pi}{3}\Gamma\int_{t_c}^t dt' \,v_w^3\left(\int_{t'}^t \frac{d\tilde{t}}{a(\tilde{t})}\right)^3 =  \frac{4\pi}{3}\Gamma\int_{t_c}^t dt' \,v_w^3\left(\int_{t'}^t \frac{d\tilde{t}}{e^{H_V \tilde{t}}}\right)^3 \sim \frac{4\pi}{3}\left(\frac{v_w}{H_V}\right)^3  \Gamma  t\,.
\end{equation}
This quantity grows arbitrarily with time, though the total volume occupied by the false vacuum $\mathcal{V}_{false} \propto a(t)^3 \mathcal{P}_{false}$ is not decreasing, since it is inflating as well.
A stronger condition, particularly useful in these scenarios, is obtained by requiring that $\mathcal{V}_{false}$ decreases with time, that is \cite{Ellis:2018mja}
\begin{equation}
\label{strongPerCond}
\frac{1}{\mathcal{V}_{false}}\frac{d}{dt}\mathcal{V}_{false} = H(T)\left(3+T \frac{dI(T)}{dT}\right) < 0\,.
\end{equation} 
This condition puts strong constraints at lower temperatures, where vacuum energy dominates the expansion and the naive condition $I(T_p) = 0.34$ is no longer enough to ensure percolation, see Fig.~\ref{fig:percolationConditions}. In this paper, for any given value of $w$, we will take $T_p$ as the minimum temperature for which all the three conditions are satisfied~\footnote{If for a given value of $w$ more than a solutions is present, we take the one at the higher temperature, since it will be reached earlier during the Universe evolution.}. In practice we find that for the $U(1)$ model under consideration the condition $I(T_p) = 0.34$ guarantees the fulfilment of all the necessary conditions for the successful percolation for  values of $w\lesssim 10^{15}$ GeV, which are the values of $w$ which can be realistically probed by present and future experiments. At last, we would like to notice that for U(1) model, the percolation and the phase transition can only occur for the values of couplings $g\gtrsim 0.5$. For the lower values of couplings, the percolation temperature drops infinitely close to zero and we have checked numerically that there will be no phase transition for the temperatures down to $\sim 10^{-15}w$. At this point, we do not expect our toy model to be valid anyway in such enormous energy range so we will consider this region of parameter space $(g<0.5)$ to be without phase transition.

\section{Bubble dynamics and GW }
\label{sec:dynamics}
Now that we have determined the conditions for a successful phase transition, we briefly discuss the dynamics of the bubbles expansion. If the Universe undergoes a very strong phase transition, the bubbles expand at supersonic speed and may reach relativistic velocities according to the strength of the phase transition \cite{Espinosa:2010hh}. Hereafter, we will assume that the bubbles are free to expand in the surrounding plasma and that they can reach relativistic velocities.
 In this regime the computation of the forces acting on the bubble wall  are simpler and  we review them in the next section, while in Sec.~\ref{sec:results_u1} we will present the resulting sensitivity of the various gravitational wave interferometer experiment for the case of the higgsed $U(1)$ model.

\subsection{The effect of friction forces on the bubbles expansion
}

Relativistic expanding bubbles of true vacuum experience a friction force on their wall acting as an inward pressure. This happens when there are particles with a phase dependent mass term, {\emph{i.e.}} particles that change their mass during the crossing of the wall between the unbroken and broken symmetry phase~\cite{Bodeker:2009qy,Espinosa:2010hh,Bodeker:2017cim}. Working in the plasma reference frame, the leading order (LO) term of the friction force $P_{{\rm{LO}}}$ turns out to be independent on the Lorentz parameter $\gamma$ associated with the bubble wall expansion. This means that relativistic bubbles can successfully expand and keep accelerating if the driving force due to the vacuum energy released during the PT overwhelms the friction force, that is $\Delta V >P_{{\rm{LO}}}$. In this case, most of the released energy is converted to accelerate the bubble wall and bubble collisions will dominate the GW signal. By explicit computation (see ~\cite{Espinosa:2010hh}), the necessary condition of successful accelerated expansion becomes 
\bea
\label{eq:lofriction}
\Delta V > P_{{\rm{LO}}}=\frac{T^2}{24}\sum_{light\to heavy}c_i N_im_i^2, 
\eea
 where $c_i=1\;(1/2)$ for bosons\;(fermions) and $N_i$ is the number of 
degrees of freedom with a phase dependent mass term \footnote{Similar effect due to the change of the mass of the scalar field (Higgs component) is loop suppressed and we ignore it.}. 
 However it was recently found that this condition of accelerated expansion is no longer true if NLO effects are considered. In particular, in the presence of massless vector bosons gaining a mass in the true vacuum phase, NLO effects due to the additional light field emissions turn out to be $\gamma$-dependent. In particular, Ref.\cite{Bodeker:2017cim} find this extra contribution to the inward pressure acting on the bubble wall to be equal to
\bea
\label{BMfric}
P_{{\rm{NLO}}}\simeq \frac{1}{16\pi^2}T^3 \gamma g^3 \Delta \phi ,
\eea
where the additional $16\pi^2$ factors comes from the phase space integration. Consequently, there is a maximum value for the $\gamma$ parameter which is given by the equilibrium conditions between the vacuum energy and the friction forces given by \cite{Ellis:2018mja}
\bea
\gamma_{eq}=\frac{\Delta V-P_{{\rm{LO}}}}{T^3 g^3 \Delta \phi /(16\pi^2)}.
\eea
The NLO friction effect will become important only for  bubbles that can reach a Lorentz factor larger than $\gamma_{eq}$ in absence of NLO friction. It then becomes necessary to find whether bubbles can reach such velocities at the moment of collision.  This can be done  by calculating the quantity $\gamma_*$, that is the Lorentz factor at the collision point without considering the NLO friction~\cite{Ellis:2018mja}. Then obviously NLO friction becomes important only $\gamma_*> \gamma_{eq}$. The Lorentz factor $\gamma_*$ can be calculated by equating the surface energy of the bubble to the gain in the potential energy:
\begin{equation}
\label{energyCons}
4\pi R^2_*\gamma_*\sigma =\frac{4}{3}\pi R^3_* (\Delta V-P_{{\rm{LO}}} )\,,
\end{equation}
which implies
\begin{equation}
\gamma_{*} =\frac{R_*(\Delta V-P_{{\rm{LO}}})}{3 \sigma}
\end{equation}
where $\sigma$ is the surface energy density of the bubble.
For $O(3)$ symmetric thin-wall bubbles with surface energy density $\sigma_{thin}$ the action takes the form \cite{Linde:1981zj}
\bea
\label{actionS3}
S_3(R)=4\pi R^2 \sigma_{thin} -\frac{4}{3}\pi R^3 \Delta V,
\eea
which can be used to estimate the surface energy density appearing in Eq.~(\ref{energyCons}).
Extremizing the action, we find the critical value of the radius
\bea
R_{thin}=\frac{2\sigma_{thin}}{\Delta V}=\l(\frac{3}{2\pi}\frac{S_3}{\Delta V}\r)^{1/3},
\eea
so that  the $\gamma_*$ depends on the radius of bubble as
\bea
\label{eq:gammatw}
\gamma_*=\frac{2}{3}\frac{R_*}{R_{thin}}\l(1-\frac{P_{{\rm{LO}}}}{\Delta V}\r)
\eea
At the time of nucleation for the models under considerations bubbles are generically thick-wall, so our approximation would look too naive. We however anticipate here that for the specific case of the higgsed $U(1)$ model
we have analyzed numerically the time evolution of the bubble solutions $\phi(r,t)$ and we typically find  the following behaviour: immediately
after the nucleation, the bubble radius $R_0$ does not change  significantly while $\phi(r=0,t)$ reaches the true minimum; 
then the bubble starts  expanding (see Appendix \ref{sec:timedep} for details).
This allows us to estimate $\gamma_*$ using the Eq.~\eqref{eq:gammatw} with a substitution
\begin{equation}
\label{eq:criticalGamma}
\gamma_*\simeq\frac{2}{3}\frac{R_*}{R_0}\l(1-\frac{P_{{\rm{LO}}}}{\Delta V}\r) \ ,
\end{equation}
where, again, we stress that this approximated equality is valid in the higgsed $U(1)$ model.

We can see that the value of the Lorentz factor $\gamma_*$ depends on the bubble radius $R$ at the moment of collision which can be estimated as \cite{Cutting:2018tjt,Hindmarsh:2017gnf}
\begin{equation}
\label{eq:rad}
R_*=\l(n_{B}\r)^{-1/3},
\end{equation}
where $n_B$ is a number density of the bubbles at the moment of  collision. In turn the bubble number density at the moment of percolation can be find as
\bea
n_B=\frac{N_{B}}{a(T_p)^3 \mathcal{V}}=\frac{1}{a(T_p)^3}\int_{t_c}^{t_p} dt'\,\Gamma(t') 
a(t')^3 \mathcal{P}_{false}(t')\,,
\eea
where $\mathcal{V}$ is a comoving volume.
Since $\mathcal{P}_{false}(t')$ changes only between $[0.7,1]$, see Eq.~\eqref{eq:percolation}, we can safely ignore it inside the integral. Then, using the time-temperature relation $dt/dT = -\left(TH(T)\right)^{-1}$ and $a(T) T= const$, we get
\begin{equation}
\label{eq:density}
n_{B}= \int_{T}^{T_c} \frac{d T}{T}\frac{\Gamma(T)}{ H(T)}\left(\frac{T_p}{T}\right)^3
\end{equation}
where $n_B$ is the number density of the bubbles at the percolation temperature. Using Eq.~\eqref{eq:density} and Eq.~\eqref{eq:rad} we can therefore  directly  compute   $R_*$
once we know $\Gamma(T)$ (note that this is exactly the quantity reported in the simulations \cite{Cutting:2018tjt,Hindmarsh:2017gnf}). In a non-expanding Universe case this quantity can be related to the $\beta$ parameter of the phase transition \cite{Enqvist:1991xw}
\begin{equation}
\label{eq:beta}
\beta=\frac{(8\pi)^{1/3}{ v_w}}{R_*}.
\end{equation}
\bfc
\includegraphics[scale=0.8]{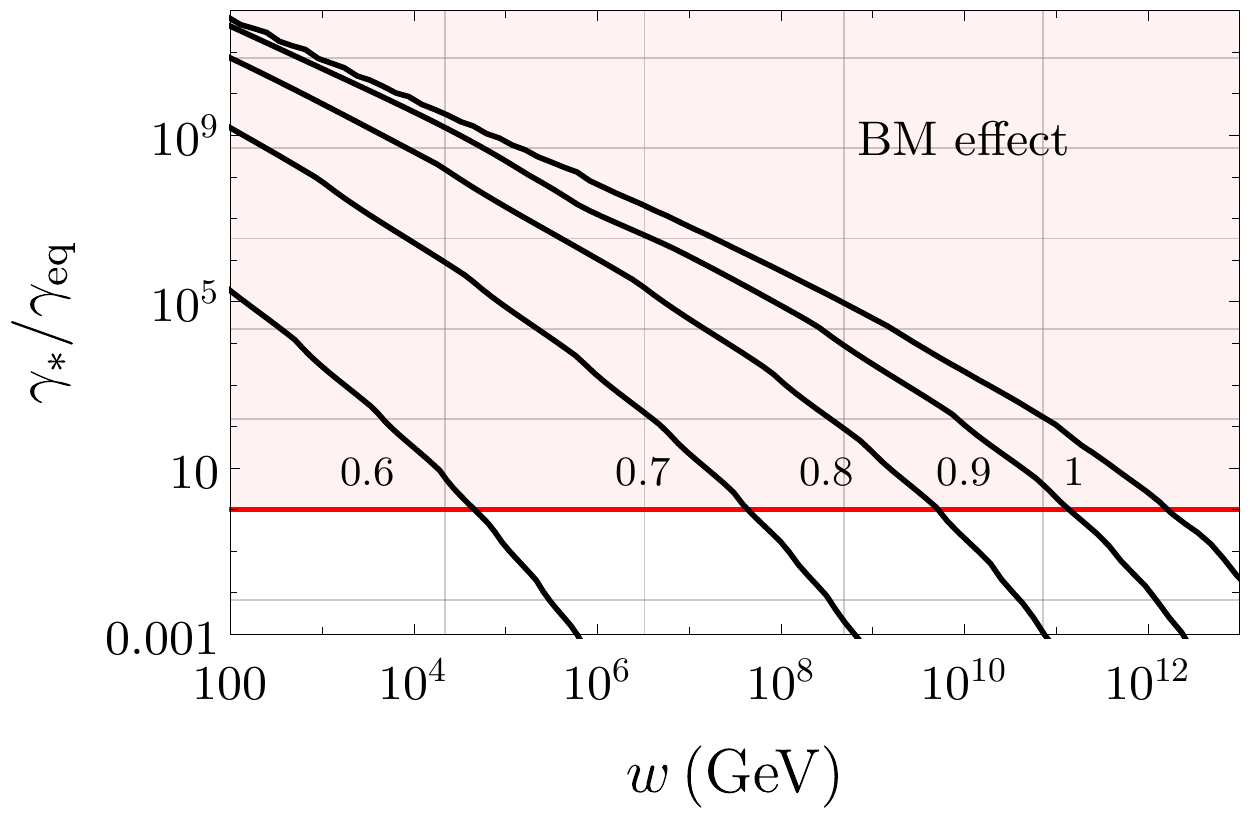}
\caption{{\it Ratio $\gamma_{*}/\gamma_{eq}$\, for different values of the $U(1)$ gauge coupling $g$ in function of the $w$ scale of the model. The shaded area represent the region where NLO friction effects are important. 
}}
\label{fig:gratio}
\efc
Combining all together we can estimate the Lorentz factors $\gamma_*$ and $\gamma_{eq}$.   Again, for the case of  the higgsed $U(1)$ model, we then show in Fig.~\ref{fig:gratio}  the ratio $\gamma_*/\gamma_{eq}$ for different values of  the $U(1)$ gauge coupling strength $g$. In particular we see that  NLO friction term effects are more important 
for smaller values of the $w$ scale of the model for a fixed value of the gauge coupling $g$.

\subsection{Energy distribution}
\label{sec:spectra}

In order to compute the stochastic GW spectrum arising from the FOPT  we need to know how the available total energy gets distributed between the bubble wall, \emph{i.e.} the scalar field contribution, and the surrounding plasma, \emph{i.e.} the sound waves and plasma turbulence contributions. The various contributions can be estimated as follows~\cite{Ellis:2019oqb,Ellis:2018mja,Fairbairn:2019xog}.

If the bubble has reached the equilibrium value $\gamma_{eq}$, then the wall will keep expanding at constant velocity. Consequently, while the energy stored in the wall will keep growing as $R^2$, the total available energy will grow as $R^3$, {\emph{i.e.}} the volume of the expanding bubble, making the scalar field contribution to the GW spectrum completely negligible if the expansions still lasts for a sufficient amount of time. In the opposite regime the wall is still accelerating at the time of collision and the contribution from the scalar field contribution can be important.
More concretely there are two scenarios:

\bit
\item \underline{$\gamma_{eq}> \gamma_*$: The equilibrium value for $\gamma$ is never reached:}

In this case the NLO friction term is not sufficient to prevent a runaway. The bubbles  never reach a terminal velocity and they accelerate until collision. The energy fraction that goes into the wall and fluid motions can be estimated from the energy conservation condition and they are given by {\color{blue} }
\bea
\label{eq:geqgstar1}
k_{wall}=1-\frac{\alpha_{\infty}}{\alpha},~~~~
k_{fluid}=\frac{\alpha_{\infty}}{\alpha}
\eea
where 
\bea
\label{eq:alpha}
\alpha=\frac{\Delta V(T_p)}{\rho_{R}} \ ,~~~~\alpha_\infty=\frac{P_{{\rm{LO}}}}{\rho_{R}} \eea
and where $\Delta V$ is defined in Eq.~\eqref{eq:deltaVdv}.
Note that only a part of $k_{fluid}$ will contribute to the GW spectrum from sound waves, see Eq.~\eqref{eq:ksw}.
\item \underline{$\gamma_{eq} < \gamma_*$: The equilibrium value for $\gamma$ is reached before collision:}

In this case the bubbles reach a terminal velocity. The energy fraction that goes into the wall motion can again be estimated by the energy conservation condition 
\be
\label{eq:geqgstar2}
k_{wall}=\frac{E^{wall}_{max}}{E_{total}}=\frac{4 \pi R^2 \gamma_{eq} \sigma }{4/3 \pi R^3 \Delta V}=\frac{\gamma_{eq}}{\gamma_*}\frac{\Delta V- P_{{\rm{LO}}}}{\Delta V}=\frac{\gamma_{eq}}{\gamma_*}\l(1-\frac{\alpha_{\infty}}{\alpha}\r)\\
\ee
while the fraction that goes into the fluid motion is
\be
k_{fluid}= 1- k_{wall}.
\ee
\eit
We then notice that if $\gamma_{eq}\ll \gamma_*$ the contribution from the walls collision to the GW spectrum is effectively turned off.

\subsection{GW from various contributions}
\label{sec:GW_spectra}

We can now review the  expressions for the various contributions to the stochastic GW background signal, that we recall is given by three different contributions: a scalar field contribution arising from the collision of expanding bubbles, a contribution from sound waves in the plasma and a contribution from the turbulent motion of the plasma itself.
These contributions all depend from the temperature after the phase transition. This is in generally different from the temperature $T_p$ at which the phase transition happens and is generally higher due to some amount of reheating (only a fraction of the total energy goes into the GW signal).
This temperature can be estimated from the energy conservation condition
\be
\label{eq:Treh}
(1-\Omega_{GW})\l(\Delta V +\rho_{R}\rvert_{T=T_p}\r)=\rho_{rad}|_{T=T_{reh}}
\ee
hence we have
\be
T_{reh}=T_p\l[(1+\alpha) (1-\Omega_{GW})\r]^{1/4}\simeq T_p\l[(1+\alpha)\r]^{1/4}
\ee
where  we have assumed the same number of degrees of freedom in the plasma at the reheating and percolation temperatures $T_{reh}$ and $T_p$ and in the last step we have further assumed  $\Omega_{GW}\ll 1$ , which is always the case since the GW emission is a Planck mass suppressed process.

\subsubsection{Scalar field contribution}

The fit to the latest lattice simulations  predict the following signal for the contribution directly arising from the bubble wall collisions~\cite{Cutting:2018tjt} 
\be
\frac{d \Omega_{\phi}h^2 }{d\ln k}=4.7 \times 10^{-8} \l(\frac{100}{g_*}\r)^{\frac{1}{3}}
\l({ H_{reh}} { R_*}\r)^2\l( \frac{k_{wall} \alpha}{1+\alpha}\r)^2S_{wall}(f,\tilde f_\phi)
\ee
where $g_*$ indicates the number of relativistic degrees of freedom, ${ H_{reh}}$
 and ${ R_*}$ the Hubble parameter and the bubble radius, see Eq.~\eqref{eq:rad}, evaluated at the { reheating} temperature, $k_{wall}$ is energy fraction defined in Eqs.~\eqref{eq:geqgstar1}  and \eqref{eq:geqgstar2} and $S(f,\tilde f)$ is a frequency broken power law
 \be
S_{wall}(f,\tilde f)=\frac{(a+b)^c \tilde f^b f^a}{(b \tilde f^{(a+b)/c }  +a  f^{(a+b)/c} )^c} \quad \rm{with}\quad a=3,~~b=1.51,~~c=2.18
\ee
with peak frequency
\be
\label{eq:peak_phi}
\tilde f_\phi = 16.5 \times 10^{-6}  \l(\frac{T_{reh}}{100}\r)\l(\frac{g_*}{100}\r)^{\frac{1}{6}}\l(\frac{3.2 }{2 \pi { R_*}} \frac{1} {{ H_{reh}}}\r){\rm Hz}.
\ee
In particular we can see that at low frequencies large wavelength of the signal obeys a scaling $\propto f^3$, as expected by the causality considerations~ \cite{Caprini:2009fx}.

\subsubsection{Sound wave contribution}
We use the results of Ref.~\cite{Hindmarsh:2017gnf}
\label{sec:soundwave}
 which have been obtained using a mixture of lattice and hydrodynamic simulation that predict
\be
\frac{d\Omega_{sw} h^2}{d \ln f} ={\blu 7.28} \times 10^{-5}\left( \frac{100}{g_*} \right)^{\frac{1}{3}} \l(\frac{k_{sw}\alpha}{1+\alpha}\r)^2
 \l({ H_{reh}} { R_*} \r) {\min}[1,\frac{{ H_{reh}} { R_*}}{U_f}] \;\tilde \Omega_{GW} S_{SW}(f,\tilde f_{sw}) \label{eq:sw}
 \ee
 with power law
 \be
C(s)=\left(\frac{f}{\tilde f}\right)^3\l(\frac{7}{4+3 (f/\tilde f)^2}\r)^{\frac{7}{2}}
 \ee
 and peak frequency
 \be
  \label{eq:peak_sw}
\tilde f_{sw} \simeq 26 \times 10^{-6} \l(\frac{1}{{ H_{reh}} { R_*}}\r)
\l(\frac{z_p}{10}\r)\l(\frac{T_{reh}}{100\;{\rm GeV}}\r)\l(\frac{g_*}{100}\r)^{\frac{1}{6}}\rm{ Hz}
\ee
evaluated with $z_p=6.7$ and $\tilde \Omega_{GW}=0.12$ and where $U_f$ is a root mean square velocity of the plasma motion.
The calculation performed in Ref. \cite{Hindmarsh:2017gnf} has been done  only in the regime ${ H_{reh}} { R_*}/U_{f}>1$ and the authors have found that the duration of the GW source is roughly $\sim 1/{ H_{reh}}$. However in the opposite case, where ${ H_{reh}} { R_*}/U_{f}<1$, we can expect that the duration of the source scales roughly as $\sim { R_*}/U_f$, so that 
a factor $\min[1,{ H_{reh}} { R_*}/U_{f}]$ appears in Eq.~\eqref{eq:sw}, see also \cite{Ellis:2018mja}. On top of this we know that the turbulent motion  will develop in a timescale $\sim { R_*}/U_f$ which again motivates  the factor 
$\min[1,{ H_{reh}} { R_*}/U_{f}] $. 
Finally, the efficiency factor $k_{sw}$ was calculated for various bubble wall velocities in~\cite{Espinosa:2010hh} and for the case of relativistic walls it was found to be
\bea
\label{eq:ksw}
k_{sw}=\l\{ \baa{c} \gamma_*>\gamma_{eq} ,~~~ \l[1-\frac{\gamma_{eq}}{\gamma_*}\l(1-\frac{\alpha_{\infty}}{\alpha}\r)\r]f(\alpha)\simeq f(\alpha)\\
\gamma_*<\gamma_{eq} ,~~~ \frac{\alpha_{\infty}}{\alpha} f(\alpha_{\infty})
\eaa\r.
\eea
with
\be
f(\alpha)\sim \frac{\alpha}{0.73+0.083 \sqrt{\alpha}+\alpha}\quad {\rm{and}}\quad U_f^2=\frac{3}{4}\frac{k_{sw}\alpha}{1+\alpha)}
\ee
where the factor $[1-\frac{\gamma_{eq}}{\gamma_*}\l(1-\frac{\alpha_{\infty}}{\alpha}\r) ]$ comes from Eq.~\eqref{eq:geqgstar2}.
 In the case of non-relativistic bubble expansions, it becomes necessary to calculate the velocity of the bubble wall. This requires to solve complicated transport equations  (see {\emph{e.g.}}~\cite{Moore:1995si,Moore:1995ua,Dorsch:2018pat,
Konstandin:2014zta,Leitao:2014pda}), which is however a task beyond the scope of this paper.  By using 
the results of~\cite{Dorsch:2018pat} we note however that for the typical strength of the FOPT obtained in this model,  {{$(\langle \phi \rangle / T)|_{T=T_c} \sim 3-10$}}, we have $v_w\gtrsim 0.1$. We thus take in the following $v_w=0.1$ as a conservative estimate for the bubble wall velocity.
In this case 
the efficiency factor becomes \cite{Espinosa:2010hh} 
\be
\label{eq:nonrel}
k_{sw}\sim \frac{v_w^{6/5} 6.9 \alpha}{1.36-0.037\alpha^{1/2}+\alpha},~~\hbox{for}~~ v_w< 0.1 .
\ee

\subsubsection{Magneto- and hydrodynamic-turbulence contribution}\label{sec:turb_contr}

As mentioned in the Introduction we will neglect in the presentation of our results the contribution from the magneto and hydrodynamic turbulence effects, in view of the latest recommendation of the LISA Cosmology Working Group~\cite{Caprini:2019egz}, due to the uncertainties associated with their calculation. However, since we will at least qualitatively comment on how the inclusion of this contribution might affect our results, we here report the parametrization of this source of GW background~\cite{Caprini:2015zlo}

The contribution from the magneto and hydrodynamic turbulence can be parametrized as follows 
\be
\Omega _{turb}h^2= 1.14 \times 10^{-4}({ H_{reh}} { R_*})\l(\frac{k_{turb} \alpha}{1+\alpha}\r)^{\frac{3}{2}}\l(\frac{100}{g_*}\r)S_{turb}(f,\tilde f_{turb})
\ee
with power law
\be
S_{turb}= \frac{(f/\tilde f )^3}{\l(1+f/\tilde f \r)^{\frac{11}{3}}(1+8\pi f/h_*)}
\ee
peak frequency
\be
 \label{eq:peak_turb}
\tilde f_{turb}=7.9 \times 10^{-5} \left(\frac{1}{{ H_{reh}} { R_*}}\right)\left(\frac{T_{reh}}{100 \rm{GeV}}\right)\l(\frac{g_*}{100}\r)^{\frac{1}{6}} {\rm Hz}
\ee
and
\be
h_*=16.5 \times 10^{-6} \left(\frac{T_{reh}}{100 \rm{GeV}}\right)\l(\frac{g_*}{100}\r)^{1/6} {\rm Hz}.
\ee
where we have expressed the relations of~\cite{Caprini:2015zlo} in function of ${ R_*}$ by using $\beta=(8\pi)^{1/3}v_w/{ R_*}$. 

\subsection{Experimental reach on the model parameter space}
\label{sec:results_u1}

The sensitivity of the various gravitational wave interferometer experiments in detecting the stochastic GW signal is controlled by the time integrated signal-to-noise ratio \footnote{  For single-detector experiments such as LISA, the definition of SNR is reduced by a factor of 2, see  \cite{Breitbach:2018ddu} for details. }
\begin{equation}
\label{thrCond1}
\rho^2= 2t_{obs}\int_{f_\text{min}}^{f_\text{max}}d f \left(\frac{h^2 \Omega_{GW}(f)}{h^2 \Omega_{noise}(f)}\right)^2,
\end{equation}
where $\Omega_{noise}(f)$ is the instrumental noise of a give experiment, sensitive to the frequency range $(f_{min},f_{max})$, and $t_{obs}$ is the observational time of the experiment. The signal is assumed to be observable if $\rho$ is greater than some threshold value $\rho\gtrsim \rho_{thresh}$, generically taken to to be $\rho_{thresh}=10$. We refer to App.~\ref{sec:gwexp} for the details on derivation of the experimental sensitivity curves and the references to the various experiments.

\begin{figure}[t!]
\centering
\includegraphics[scale=0.65]{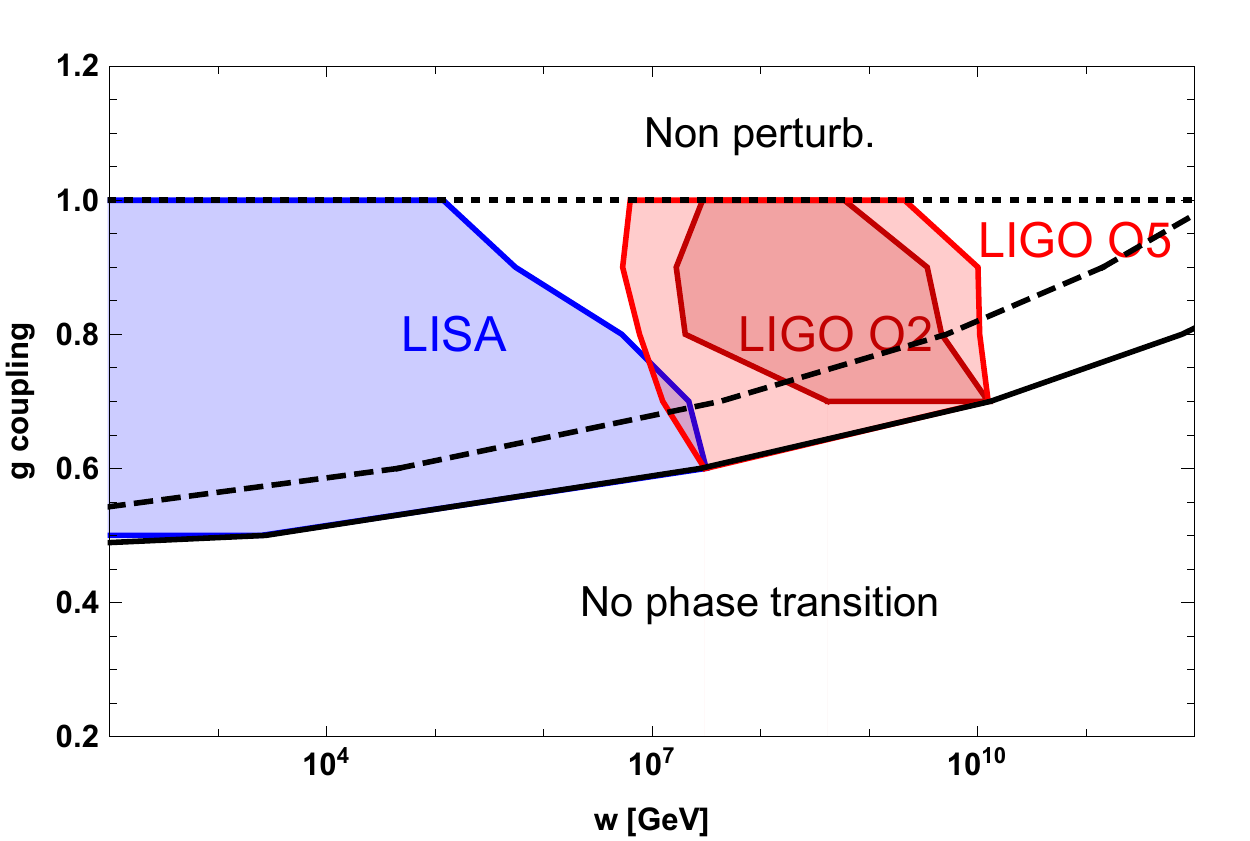}
\includegraphics[scale=0.65]{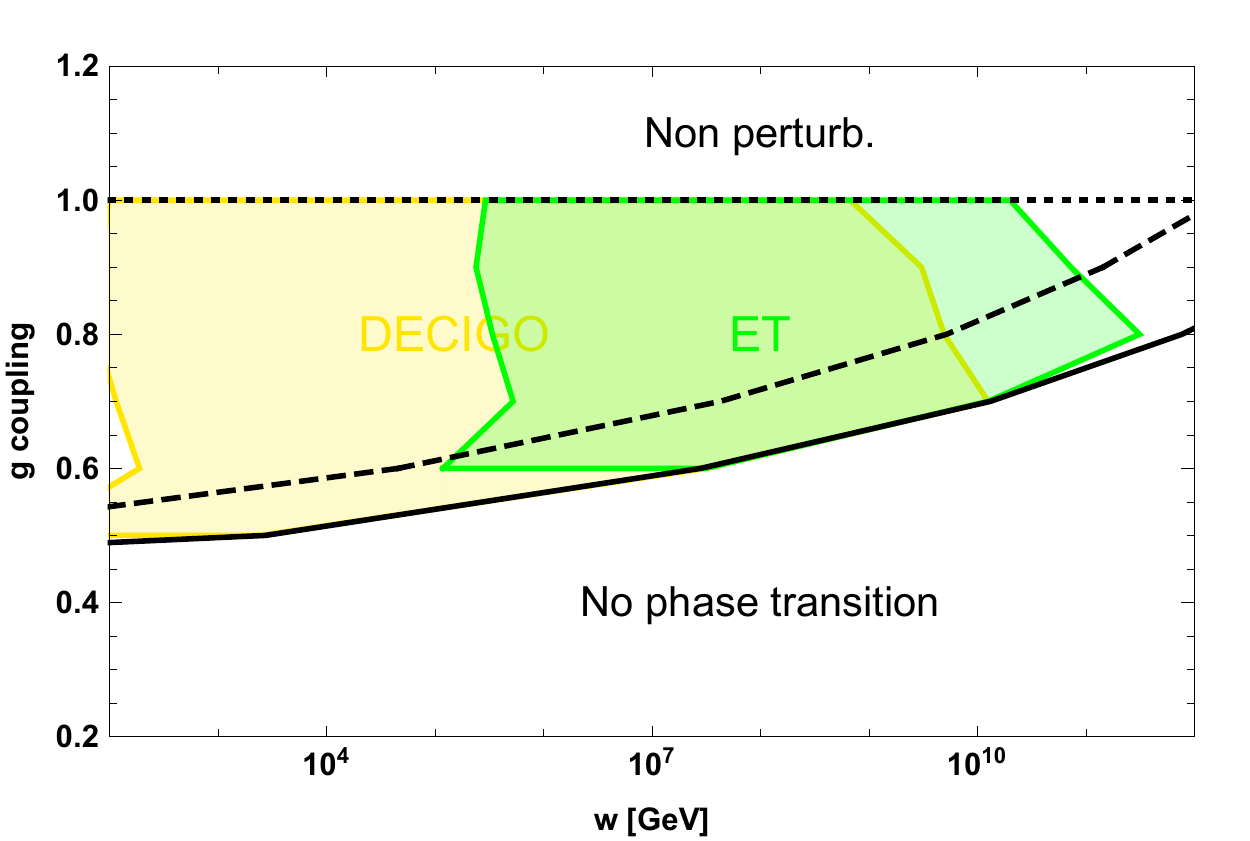}
\includegraphics[scale=0.65]{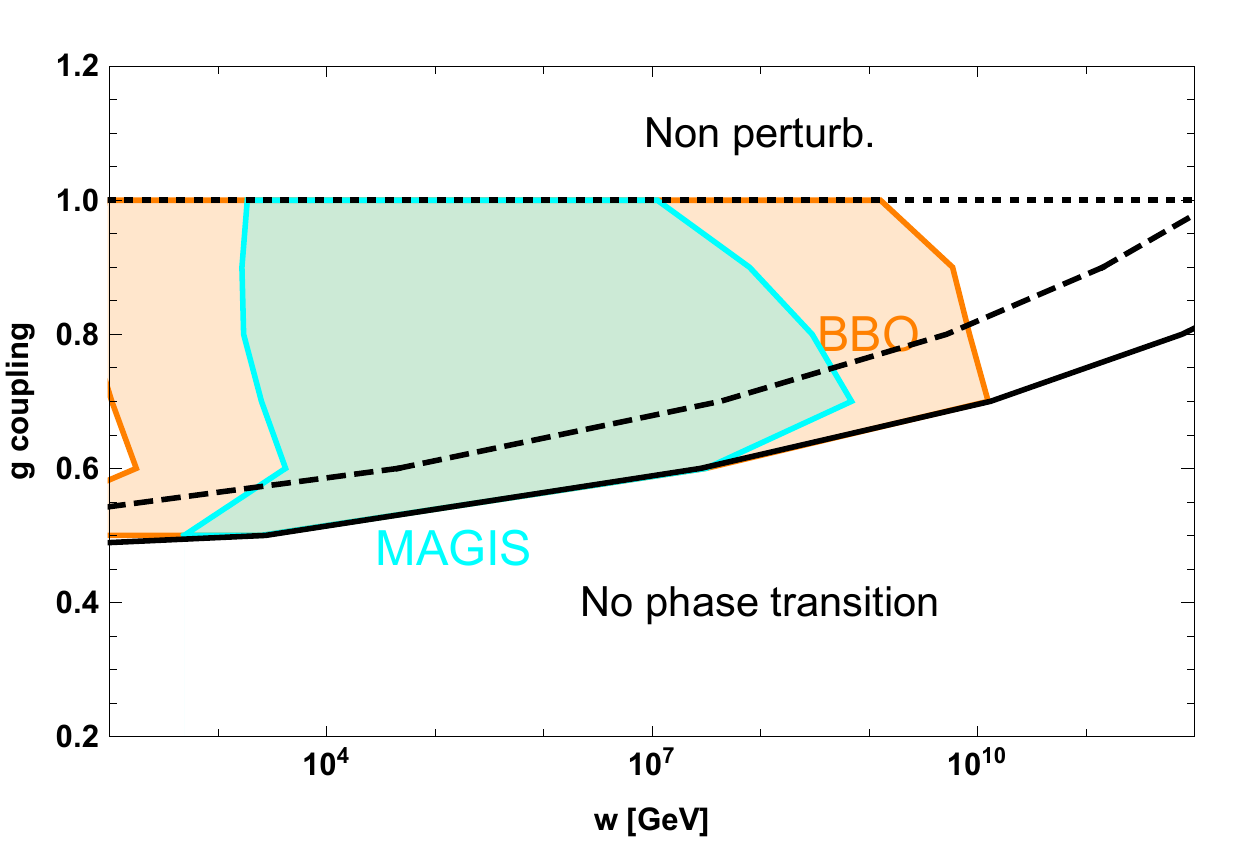}
\caption{{\it Reach of  the various experiments in the parameter space of the $U(1)$ model. In the region below the lower black solid curve there is no phase transition. Between the lower black curve and the dashed black curve the NLO friction effects are irrelevant.  Above the black dotted line, where $g>1$, perturbative calculations cannot be safely trusted as mentioned in Sec.~\ref{sec:model_1}.
Colored contours indicate the reach of the various experiments.}
 \label{fig:boundsmodel1} 
}
\end{figure}

For the case of the higgsed $U(1)$ model the expression for the LO pressure of Eq.~\eqref{eq:lofriction} reads now
\be
P_{LO}=3\frac{T^2}{24}  g^2 \Delta \phi^2 \ ,
\ee
and we present our results in Fig.~\ref{fig:boundsmodel1} in function of the model scale $w$ and $U(1)$ gauge coupling $g$.

 The left and right panels correspond to the sensitivity of different present and future experiments, reported separately for clarity. 
In both panels above the black dashed line the NLO friction effect is relevant, a region of parameter space not discussed in~\cite{Breitbach:2018ddu} where the same models is studied, while~\cite{Bian:2019szo}, where  a classically conformal $U(1)_{B-L}$ model is analyzed, without however discussing the friction effects.
 
We also indicate the region where the PT cannot occur, where the success of the PT is guaranteed by the conditions explained in Sec.~\ref{sec:percolation}, while above the black dotted line, where $g>1$, perturbative calculations cannot be safely trusted as mentioned in Sec.~\ref{sec:model_1}.
Another source of uncertainty is that for the model under consideration we are never in the condition where ${ H_{reh}} R_*/U_f <1$, \emph{i.e.} where the calculation of~\cite{Hindmarsh:2017gnf} is completely reliable, which forces us to extrapolate the signal outside the simulation range of validity.

All together we see that the current phase of LIGO is currently sensitive to scales between $w=10^7-10^9\;$GeV for couplings values between $g=0.7-1$ while its future upgrade could test up to $w=10^{10}\;$GeV.  On the other side {  LISA will complement these results by  testing  lower  scales in the same coupling value range }while other future planned experiment such as BBO, DECIGO, MAGIS and ET will greatly enlarge the reach onto the model parameter space and will be able to test models with a dark phase transition up to an energy scale of $\sim 10^{12}\;$GeV.  
Interestingly, in the low $w$ side of the bounds, we can see a characteristic change of exclusion lines. This is evident for example 
for the case of the ET experiment with $w=5\times 10^5$ GeV around $g\sim 0.6$, which is the region where the BM effect becomes important. This effect is related to the fact that below the dashed line the GW signal is effectively dominated by bubble collision. At high frequency this contribution is decaying more slowly than the contribution arising from sound waves, $\sim f^{-3/2}$ against $\sim f^{-4}$, thus increasing the sensitivity of the experiments.

%%%%%%%%%%%%%%%%%
%% MODEL 2 %%%%%%%%%
%%%%%%%%%%%%%%%%%

\section{Two scalars model}
\label{sec:model_2}

In the previous section we have shown that the NLO friction effect has a relevant impact in the higgsed $U(1)$ model parameter space available at present and future interferometer experiments, see Fig.~\ref{fig:boundsmodel1}. 
As reviewed, this effect strongly reduces the efficiency factor relative to the GW spectrum produced by the collision of the true vacuum bubbles. 
In view of this, it is then interesting to study whether there are different models that have a scalar potential with the {\emph{same shape}} as the one of the higgsed $U(1)$ one, where however the NLO friction effect is absent, thus producing a {\emph{very different}} GW signal. As shown in Sec.~\ref{sec:GW_spectra}, the amplitudes and peak frequencies of the various sources of the GW spectra are controlled by the following parameters
\be
\alpha/(1+\alpha), \quad
 T_{reh}\sim T_{p}(1+\alpha)^{1/4},\quad
 { H_{reh}} { R_*}.
 \ee
If these quantities do match for different models around the percolation temperature $T_p$, the various contributions to the total GW signal will have the same peak frequencies and shapes. However the efficiencies factors for the different contributions can vary, since they depend on the importance of the friction effect in the different models, thus determining different relative amplitude and therefore a different spectrum for the {\emph{total}} signal.
In particular the peak frequencies are roughly in the following proportion, see Eqs.~\eqref{eq:peak_phi}, \eqref{eq:peak_sw} and \eqref{eq:peak_turb},
\be
f_{\phi}:f_{sw}:f_{turb}\simeq 1:2:9,
\ee
while the high-frequency behavior of the spectra scales approximately as
\be
\label{eq:freq_prop}
\Omega_{\phi} \sim f^{-3/2}, \qquad \Omega_{sw} \sim f^{-4}, \qquad \Omega_{turb} \sim f^{-5/3},
\ee
so that looking at the peak dependence of the total signal could provide information relatively to the different contributions to the GW spectra in various models.

Since the friction effect is relevant for theories where gauge bosons acquire a mass after the spontaneous breaking of a symmetry
\footnote{So far  $\gamma$ dependent  friction effects 
have been found only due to the soft particle production \cite{Bodeker:2017cim} and in this case only  in the theories 
 where the vector boson get masses during the phase transition.
 Of course this does not guarantee that some other higher order effect can lead to another contribution to the friction force proportional to 
$\gamma$.},
 we can consider as the most minimal scenario  a theory with just a single scalar field  $\phi$ with a tree-level barrier induced by a cubic term $\sim k \phi^3$. In this case however the field never undergoes a true transition form the false to the true vacuum and no first oder phase transition can occur, see {\emph{e.g.}}~\cite{Breitbach:2018ddu}.
The situation is however different if one adds a second auxiliary scalar singlet, whose role is to mimic the effect of the vector boson of the Higgsed $U(1)$ case in order to radiatively induce a barrier between the false a true vacuum and produce a FOPT.

As a toy example, lets consider a model with two scalar fields, $\phi$ and $\eta$, which are even and odd under a ${\mathbb Z}_2$ symmetry respectively. The general Lagrangian for this theory is given by
\bea
\label{eq:scalar}
{\cal L}=\frac{1}{2}(\d_\mu \phi)^2+\frac{1}{2}(\d_\mu \eta)^2- \frac{k}{3} \phi^3-k_{\phi \eta}\phi \eta^2-\frac{\lambda_\phi}{4} \phi^4-\frac{\lambda_\eta}{4} \eta^4- \frac{\lambda_{\phi \eta}}{2} \phi^2 \eta^2- \frac{\mu^2_{\phi}}{2} \phi^2- \frac{\mu_\eta^2}{2}\eta^2. 
\eea
The ${\mathbb Z}_2$ symmetry forces the tunnelling to happen along the $\phi$ direction, reducing the problem of finding the two dimensional bounce solution  to an effective one dimensional problem.

We can further simplify the study by considering only the necessary ingredients to produce a one-loop effective potential with a similar shape of the higgsed $U(1)$ case. Practically, one only needs to introduce a mass term for the $\phi$ field, a quartic interaction for $\eta$, to guarantee the stability of the potential, and a quartic mixing between $\phi$ and $\eta$. All together we will study the following theory
\bea
\label{eq:scalarsimp}
{\cal L}=\frac{1}{2}(\d_\mu \phi)^2- \frac{\mu_{\phi}^2}{2} \phi^2+\frac{1}{2}(\d_\mu \eta)^2-\frac{\lambda_\eta}{4} \eta^4- \frac{\lambda_{\phi \eta}}{2} \phi^2 \eta^2.
\eea
In this case the field dependent mass plus thermal dressing read
 \be
 \begin{split}
 & m_\phi^2+\Pi^2_{\phi}=\mu_{\phi}^2+\lambda_{\phi \eta}\eta^2+\frac{\lambda_{\phi \eta} T^2}{12}\\
 & m_\eta^2+\Pi^2_{\eta}= \lambda_{\phi \eta}\phi^2+3\lambda_\eta \eta^2+T^2\l(\frac{\lambda_\eta}{4}+\frac{\lambda_{\phi \eta}}{12}\r)
 \end{split}
 \ee
and we again use the truncated full dressing procedure~\cite{Curtin:2016urg} to compute the potential:
 \begin{equation}
  V(\phi,\eta,T)=V_{CW}(m_i^2+\Pi_i^2)+V_T(m_i^2+\Pi_i^2).
 \end{equation}
 In this model the LO friction term is different with respect to the Higgsed $U(1)$ case and reads
\be
P_{{\rm{LO}}}= \frac{T^2}{24}\lambda_{\phi \eta} S^2.
\ee

From the plots of Fig.~\ref{fig:percolationConditions} we observe that percolation in the first model occurs for relatively small values of $T/w$ for $g\sim 1$. In this case it is possible to perform an approximate mapping between the two models, where the mapping is imposed by requiring approximately
\be
\begin{split}
& \langle \phi \rangle_{\rm true}^{\rm{model 1}} \simeq \langle \phi \rangle_{\rm true}^{\rm{model 2}} \\
& \Delta V_{{\rm{model 1}}} \simeq  \Delta V_{{\rm{model 2}}}\;.
\end{split}
\ee
This works because at low $T$ the temperature dependence on the pure CW part is mild (not true for $V_T$) and we can reabsorb the differences of the two models through the two renormalization scales.  Also, $\lambda_{\phi \eta }$ should scale as $\sim g^2$. 
Numerically, by imposing
\be
V^{\rm{model 1}}(x,T=0)|_{\mu_R=g w} = V^{\rm{model2}}(x,0,T=0)_{\mu_R=a \sqrt{\lambda_{\phi \eta}}w}
\ee
and solving for $a$ and $\lambda_{\phi \eta}$ we find
\be
a \sim 1.65, \quad\quad \frac{\lambda_{\phi \eta}}{g^2} \sim 1.73.\\
\ee
An approximate mapping is then obtained by imposing
\be
\begin{split}
& \mu_{R}^{\rm{model1}} = g w, \quad\quad \mu_{R}^{\rm{model2}}  = 1.65 \sqrt{\lambda_{S\eta}} w, \quad\quad  \lambda_{\phi \eta} \sim1.73 g^2.
\end{split}
\ee
\begin{figure}[t!]
\center
\includegraphics[scale=0.8]{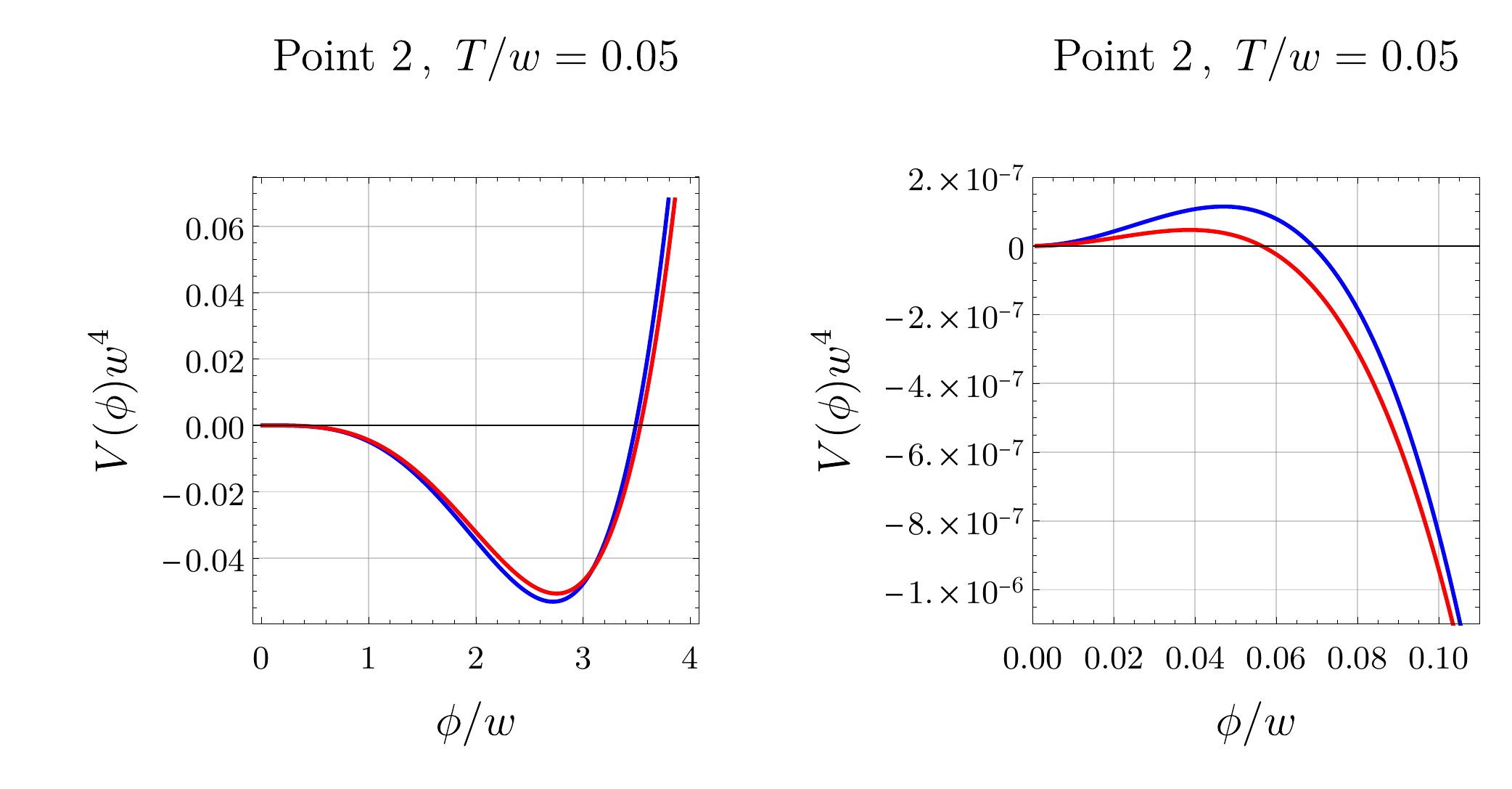}
\caption{{\it Shape of the potential for the Higgsed $U(1)$ (blue) and two singlet (red) models evaluated at $T\sim T_p$. In the right plot a zoom into the small field value is shown in order to the potential barrier between the true and false vacua.}}
\label{fig:potential_mapping}
\end{figure}
Practically, it is however necessary to also tune the $\phi$ field bare mass, in oder to adjust the potential at low field values. All together we find that it is possible to find benchmark points where the potential, percolation temperature and bubble radii at collision are the almost same~\footnote{By turning on other interactions in the model of Eq.~\eqref{eq:scalarsimp} one can obtain an even more precise matching. However the reference points mentioned in the text are sufficient to illustrate our point.}. We report in Tab.~\ref{tab:mapping} the mapping parameters for three benchmark point which are relevant for present and future GW experiments, see Fig.~\ref{fig:boundsmodel1}, where the various quantities affecting the form of the GW spectrum are also reported. For the specific case of the benchmark point 3, we show in Fig.~\ref{fig:potential_mapping} the shape of the effective potential for the field $\phi$ in the two different models, evaluated at $T=T_p$, where the right panel show a zoomed version of the effective potential at small field values to illustrate the potential barrier separating the true from the false vacua.

\begin{figure}[t!]
\centering
\includegraphics[scale=0.48]{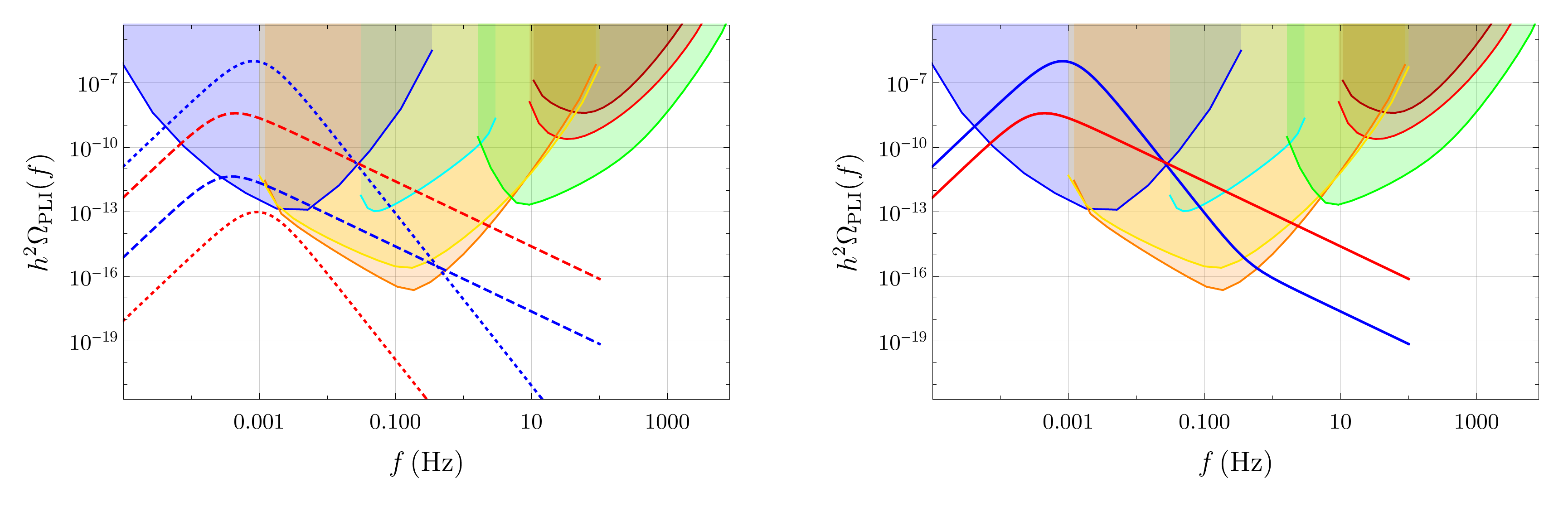}
\caption{{\it GW signal for the Higgsed $U(1)$ (blue) and the two singlet (red) models for benchmark point 1. 
The dashed and dotted  lines correspond to the bubble collision and sound waves  contribution to the GW spectra respectively. The solid lines represent the total signal.
 Also indicated are the reach of future gravitational waves experiments, see Fig.~(\ref{fig:intro}) or App.~(\ref{sec:gwexp}).}}
\label{fig:shape_diff}
\end{figure}

\begin{figure}[t!]
\centering
\includegraphics[scale=0.48]{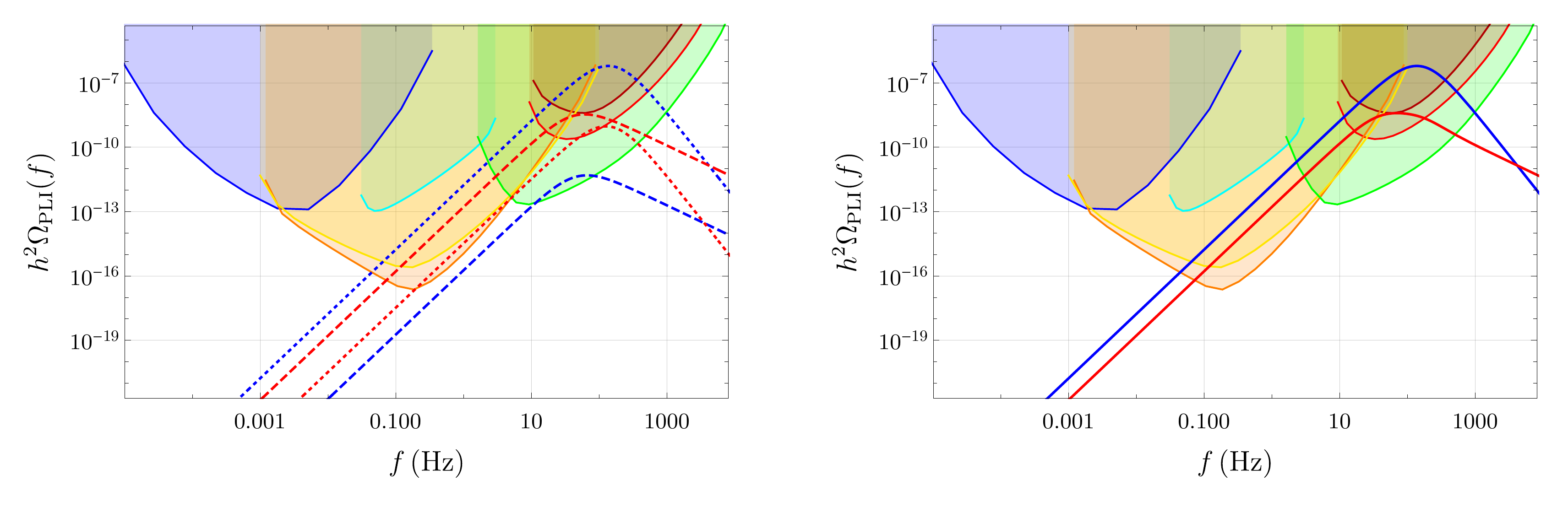}
\caption{{\it  Same as  Fig .\ref{fig:shape_diff} before for the reference point 2}}
\label{fig:shape_diff1}
\end{figure}

We then show in Fig.~\ref{fig:shape_diff} and Fig.~\ref{fig:shape_diff1} the different contributions to the GW spectra for the two models, together with the total expected signal. We can see that in the higgsed $U(1)$ case the contribution  from the bubble collisions to the total spectra is completely irrelevant, and the total shape exhibits a peak at a frequency $f \simeq 10^{-3}\;$Hz, fully determined by the sound wave contribution. On top of it at higher frequencies there is a shoulder  caused by the 
bubble collision effects  
 which we recall switches off slower than the sound wave contribution at high $f$, see Eq.~\eqref{eq:freq_prop}.  This features are completely lost in the two scalar case model, where the bubble collision  dominated the spectrum.
  The comparison between the two different total shapes can be seen in Fig.~\ref{fig:shape_diff} and Fig.~\ref{fig:shape_diff1} by comparing the solid lines.  
  We can see that for the models with the gauged symmetries the signal is 
  dominated by the sound wave contributions and for the  scalar models we have 
  only the bubble collision.    The signal from the bubble collisions drops slower at high frequencies, which offers a possibility in distinguishing two types of models. 
 The same effect leads to the  appearance  of a shoulder  in the gauged models
 at the frequencies when the bubble collision start to dominate over the sound 
 wave effects.  Note however that in order to make any precise statement, it is 
 crucial to exactly know the signal prediction. As we have shown in 
 Sec.~\ref{sec:GW_spectra}, we are however often in the region where the 
 calculation of the signal is not completely reliable. We then do not make 
 any quantitative statement regarding the model differentiation and simply 
 comment that is in principle possible to discriminate among different underlying 
 theories where friction effects are or are not relevant, by focusing on the 
 spectral shape of the GW signal.
 
 Finally, we would like to comment on the possibile impact of the inclusion of the turbulence effects in the total GW signal. As reported in Eq.~\eqref{eq:peak_turb} the peak frequency of the turbulent contribution is higher than the one of the bubble collision and sound wave contributions. Moreover, the turbulent contributions switches of at higher frequencies slower than the sound waves one,  see Eq.~\eqref{eq:freq_prop}. Thus, there could be an additional shoulder effect at higher frequencies associated with this source. However, in view of the uncertainties associated with the calculation of the GW spectra from magneto and hydrodynamic turbulence mentioned in Sec.~\ref{sec:turb_contr}, we once again do not make  any quantitative statement on the possibility of exploiting this feature to discriminate among various theories.

\begin{table}[h!]
 \begin{center}
     \begin{tabular}{c | c || c || c }
\multicolumn{2}{c||}{}      				  	& Point 1 				& Point 2 \\	
 \hline
 \hline
\multirow{4}{*}{{\textbf{Model 1}}}   &  $g$   			   		& 0.6			& 0.8\\
  				       &  $\mu_R$ 					& $gw$ 				& $gw$\\
  				       &  $\kappa_{\phi}$ 			&  			$0.03$		& $0.04$\\
  				       &  $\kappa_{sw}$ 		  					& 0.97 & 0.96\\				       				       
\hline				     
\multirow{6}{*}{{\textbf{Model 2}}}   &  $\mu_\phi^2$   		 	& 
 $6\times  2\times 10^{-8}w^2 $ 		& 0.00002$w^2$\\
				       &  $\lambda_{\phi \eta}$ 	& 1.6$g^2$  			& 1.67$g^2$ \\
				       &  $\lambda_\eta$ 				& 0.01  				& 0.01 \\
				       &  $\mu_R$ 		 & $1.71\sqrt{\lambda_{\eta\phi}}w$ & $1.65\sqrt{\lambda_{\eta\phi}}w$ \\		
  				       &  $\kappa_{\phi}$ 		&  		 		$\sim 1	$		& $\sim$0.98\\
  				       &  $\kappa_{sw}$ 			&  			$\sim    2\times 10^{-4} $		& $\sim 0.02$\\						       		       
\hline	
				 \multicolumn{2}{c||}{$w$}  & $10^{4}\;$GeV & $ 10^9\;$GeV \\		
	  \multicolumn{2}{c||}{$\alpha/(1+\alpha)$} &  1 & $\sim 1$\\	
	  \multicolumn{2}{c||}{$\langle \phi \rangle/w$} & $\sim2.7 $ & $\sim 2.7 $\\
	  \multicolumn{2}{c||}{$\Delta V/w^4$}  & $\sim 0.017  $ & $\sim 0.05$\\
	
			 \multicolumn{2}{c||}{$T_{reh}/w$}  &$ \sim 0.15$  & $\sim 0.2$ \\	
			    \multicolumn{2}{c||}{{{$1/(H_{reh}R_*)$}}} & $\sim (3.1,3.5 )$ & $\sim (3.9,3.6)$ \\	
\hline	        
 \multicolumn{2}{c||}{$\alpha_\infty$}  &($1.2\times 10^4,280 $)  &$\sim( 48,3.2)$ \\		
 \multicolumn{2}{c||}{$T_{p}/w$}  &(0.0009,0.0045)  &$\sim(0.02,0.06)$ \\		
 \multicolumn{2}{c||}{$\alpha$} & $(7.5\times 10^8,1.2\times 10^{6} )$ &$\sim(1.2\times 10^3,150)$ \\		
     \end{tabular}
 \end{center}
 \caption{{\it Input parameters for the two considered models providing an approximate mapping, see main text for details. Also reported are the quantities affecting the various contribution to the GW spectrum.}}
\label{tab:mapping}
 \end{table}

\newpage

\section{Summary and Discussion}
\label{sec:concl}
Gravitational wave astronomy is experiencing an enormous rise due to the recent 
discoveries of GW transients arising from black holes and neutron stars mergers. On the particle physics side gravitational wave detectors can provide a unique possibility in testing FOPT in the Early Universe, which are ubiquitous in many theories beyond the SM and possibly relevant to electroweak bariogenesis. 
In particular, the LIGO experiment currently operating is sensitive to 
frequencies around $10-100\;$Hz, which correspond to scales of the FOPT in the range $10^7-10^{10}$ GeV for the models that we have investigated. Future proposed experiments would extend its reach both towards lower and higher frequencies. Interestingly, interferometers such as ET would be able to test FOPT scales of $\sim10^{12}\;$GeV, which are well beyond the reach of any imaginable future collider, thus providing an extremely important probe of phenomena that might have occurred in early phases of the Universe.

At the same time in the case of discovery of a stochastic GW signal the discussion of the inverse problem, that is determining from the GW spectrum (part of) the properties of an underlying theories, is still in its infancy.  One of the main obstacles is that the precise prediction of the various contribution to the GW signal are known only in limited ranges for the parameter determining the properties of the FOPT. Beyond these validity ranges, various extrapolations are used. For example there is no numerical calculation of the turbulence contribution to the GW spectrum while the sound wave contribution is known only for phase transitions lasting longer than an Hubble time.

 Despite these uncertainties there have been great progresses in the calculation of the bubble walls velocity at the time of collision. In particular it has been recently shown that for relativistic expanding bubble walls a $\gamma$ dependent friction effect is present whenever the underlying theory present a spontaneously broken symmetry through which gauge bosons acquire a phase dependent mass. This effect might prevent the bubbles to reach a runaway conditions, strongly reducing the contribution to the stochastic GW background arising from the collision of the walls themselves. 
 
 In view of this effect, we have studied in this paper the expected signal arising from two two models, representative of two classes that can feature a FOPT. A classically scale invariant higgsed $U(1)$ model and a two singlet scenario. The former experiences a $\gamma$ dependent friction effect, while the latter does not. In turn, while very similar potentials at the time of the phase transition can ben obtained in both models, the predicted GW signal strongly differs between the two cases.  In particular, for the higgsed $U(1)$ model the spectra exhibit a shoulder at a frequency higher than the peak one. This feature is due to the fact that the bubble collision contribution , which is strongly suppressed in regions of the parameter space where the friction effect is relevant, switches off slower that the sound waves one at high frequency. This feature is almost absent for the two scalar case since no $\gamma$ dependent friction effect is preventing a runaway condition and thus the contribution from bubble collision  dominates the GW spectrum. Interestingly, these effect are well within present and future detectors sensitivity ranges for large parts of the models parameter space. 
 While no quantitative statement can be rigorously yet made, the study of this feature is of extreme importance while investigating the inverse problem of determining the underlying theory particle content and properties in the event of the discovery of a stochastic GW background. Since FOPT in the Early Universe might arise from physics lying at an energy scale well beyond the reach of any future collider experiment, this turns out to be a crucial tasks for which much more progress is required.

\section*{Acknowledgments}
We thank Lorenzo Ubaldi for collaboration at the early stages of this project. DB thanks the Galileo Galilei Institute for theoretical physics for hospitality while part of this work was carried out. The  work of AA was in part supported by the MIUR contract 2017L5W2PT.

%%%%%%%%%%%%%%%
%%%% APPENDIX %%%%%
%%%%%%%%%%%%%%%
 
\appendix
\section{GW sensitivity curves}
\label{sec:gwexp}
According to the parameter space where FOPT can occur, the crucial question is whether the GW signal generated by the PT is detectable within the sensitivity of present and future experiments. A stochastic gravitational wave background is detectable from an experiment running for a time $t_{obs}$ if the signal-to-noise ratio~\cite{Cornish:2001bb} (see also \cite{Moore:2014lga,Breitbach:2018ddu} ) is greater then a threshold value
\begin{equation}
\label{thrCond}
\rho^2= 2t_{obs}\int_{f_\text{min}}^{f_\text{max}}d f \left(\frac{h^2 \Omega_{GW}(f)}{h^2 \Omega_{noise}(f)}\right)^2 \geq \rho_{thr}^2
\end{equation}
where $h^2\Omega_{GW}(f)$ and $h^2\Omega_{noise}(f)$ are the dimensionless spectral GW and noise energy density respectively. The first takes into account various contributions form different sources such as bubble collision, sound waves or plasma turbulences, and it is generically a broken power law in the spectrum of frequencies. In the simple assumption that the GW signal follows a single power law $\Omega_{GW}(f) = \Omega_{b}(\bar{f})\left(f/\bar{f}\right)^b$ over all the frequency spectrum, the detectability condition Eq.~(\ref{thrCond}) becomes
\begin{equation}
h^2 \Omega_b(\bar{f}) > h^2 \Omega_b^{thr}(\bar{f}) \equiv \frac{\rho_{thr}}{\sqrt{2 t_{obs}}}\left[\int_{f_\text{min}}^{f_\text{max}}d f  \left( \frac{(f/\bar{f})^b}{h^2 \Omega_{noise}(f)}\right)^2\right]^{-1/2}.
\end{equation}
The choice of the reference frequency $\bar{f}$ is arbitrary and it does not affect our conclusions. For each frequency $\bar{f}$, we can define the Power-Law Integrated (PLI) sensitivity curve by maximizing over the spectral index $b$, that is
\begin{equation}
\label{defPLI}
h^2\Omega_{PLI}(\bar{f}) \equiv h^2\max_b\left[\Omega_b^{thr}(\bar{f})\right]\,,
\end{equation}
which gives the threshold value for the signal $h^2\Omega_{GW}(\bar{f})$ in order to be detectable. 
\begin{figure}[h]
\begin{center}
\includegraphics[scale=0.7]{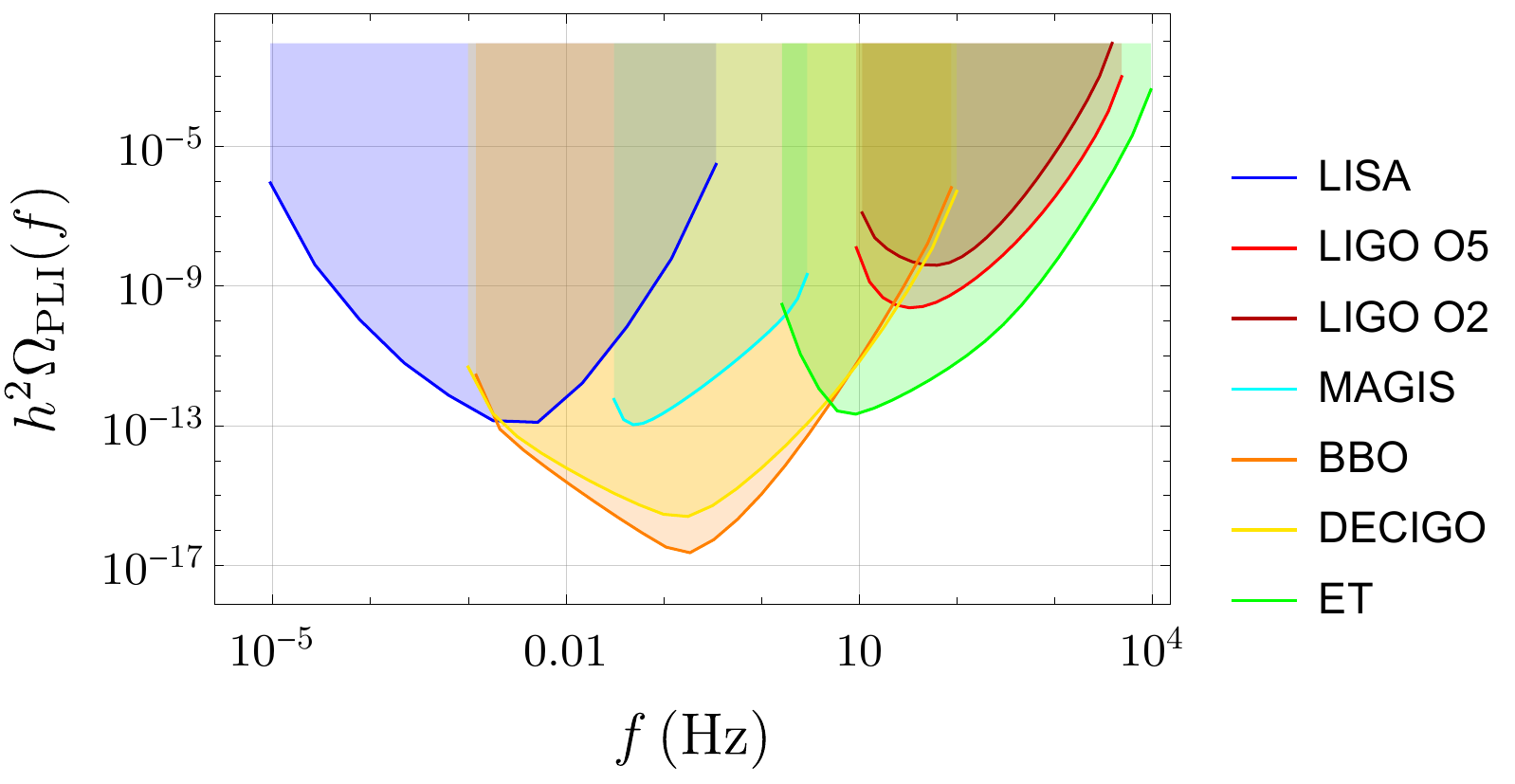}
\caption{PLI curves  for the various present and future experiments.
\label{fig:PLIcurves}
}
\end{center}
\end{figure}

Since real signals are broken power laws, we expect this interpretation to hold generically. We report in the Table~\ref{tab:expParameters} the values of $t_{obs}$ for different experiments, together with the references from which we extracted the strain noise. We plot in Fig.~(\ref{fig:PLIcurves}) the PLI curves we have derived through Eq.~(\ref{defPLI})
\begin{table*}
\caption{We take the threshold value for the signal-to-noise ratio to be $\rho_{thr}=10$ for all the experiments. For BBO and DECIGO, we convert the noise to sky-averaged strain noise, see \cite{PhysRevD.80.104009,Moore:2014lga} and references in the table. \label{tab:Parameters}}
\begin{center}
\begin{tabular}{c c c c c c c c }
&LIGO O2 & LIGO O5 & LISA & MAGIS & BBO & DECIGO & ET\\
\hline
 $t_{obs}$ (months) & 6 &20 & 48 & 60 &48 &48 &60 \\
\hline
 $h^2\Omega_{noise}$ &\cite{Aasi:2013wya}&\cite{TheLIGOScientific:2014jea} & \cite{Cornish:2018dyw} & \cite{Graham:2017pmn} &\cite{Yagi:2011yu} &\cite{Yagi:2013du} & \cite{Sathyaprakash:2012jk}
\end{tabular}
\label{tab:expParameters}
\end{center}
\end{table*}
\newpage

\section{Time dependent solution}
\label{sec:timedep}
In this appendix, we discuss the numerical solutions of propagating bubbles in vacuum. 
 We show that once the bubbles nucleate,
 their profile  quickly becomes a step function like interpolating between the true and false vacuum.  This behaviour
  justifies our estimate of the bubble Lorentz factor $\gamma_*$ in Eq.~\eqref{actionS3} evaluated at the time of percolation.

The equations of motions for $O(3)$ symmetric bubbles are given by
\begin{equation}
\label{timeDepSols}
\left(-\frac{\partial^2}{\partial t}+ \frac{\partial^2}{\partial r} + \frac{2}{r}\frac{\partial}{\partial r}\right)\phi(t,r) = \frac{\partial V(\phi)}{\partial \phi}\phi(t,r).
\end{equation}
The static solutions $\phi_{stat}(r)$ considered so far (see Fig. (\ref{fig:decayrates}) and main text) are computed by solving the boundary-value problem
\begin{equation}
\lim_{r\rightarrow+\infty}\phi_{stat}(r) = 0\,,\qquad \frac{\partial}{\partial r}\phi_{stat}(r)\Big|_{r=0}=0
\end{equation}
by means of overshooting/undershooting methods. These methods allow to determine the initial condition $\phi_{stat}(0)$ for which the solution reaches the correct asymptotic behavior with enough precision. At high temperatures $T \lesssim T_c$, the energy difference between the two vacua is small with respect to the energy barrier and  in this case\footnote{We work in units of $w=1$. } $\phi_{stat}(0) \sim \phi_1 = e$. For lower temperatures, one can solve the initial value problem $\phi_{stat}(0) = \phi_1 + \delta$ for different values of $\delta$, until $\lim_{r\rightarrow R_\infty}\phi_{stat}(r) \sim 0$ for some $R_\infty \gg 1$.

\begin{figure}[t!]
\begin{center}
\includegraphics[scale=0.42]{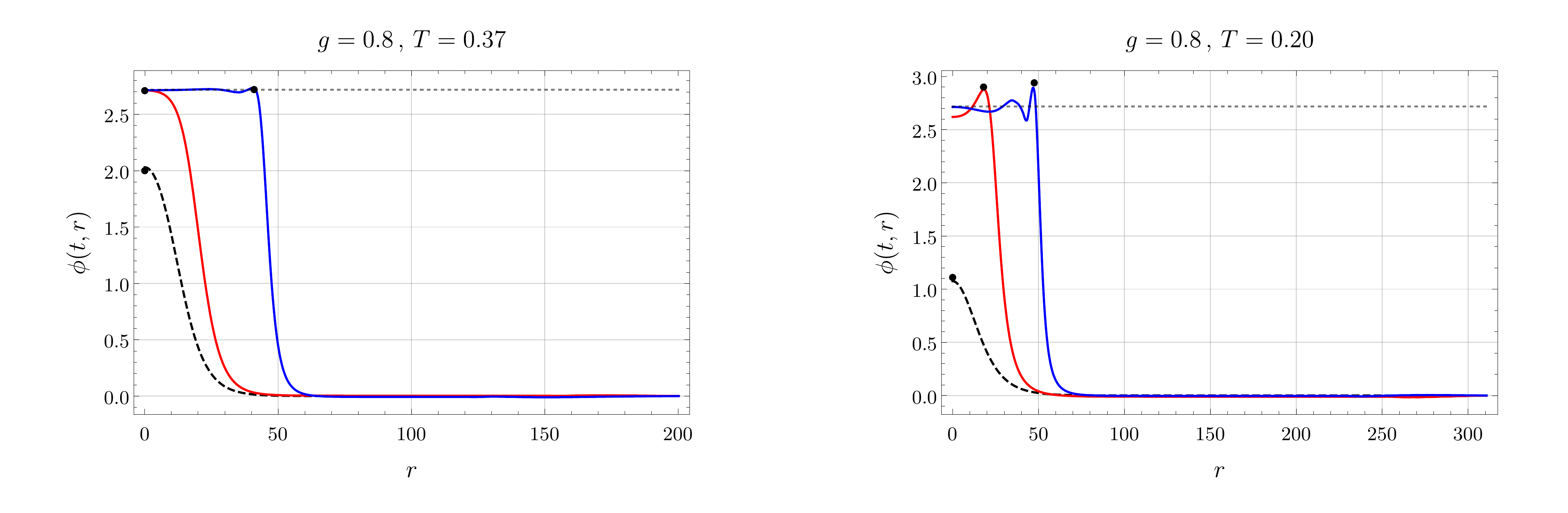}
\caption{{\it Time evolution of bubble profiles for different values of temperature and coupling $g=0.8$. We plot the solution for $t=0$ (black dashed), $t=50$ (red), $t=80$ (blue). The dotted gray line is the true vacuum value, around which $\phi(t,0)$ oscillates and eventually relaxes at late times.   }}
\label{timeEvolutionPlot}
\end{center}
\end{figure}

\begin{figure}[t!]
\begin{center}
\includegraphics[scale=0.9]{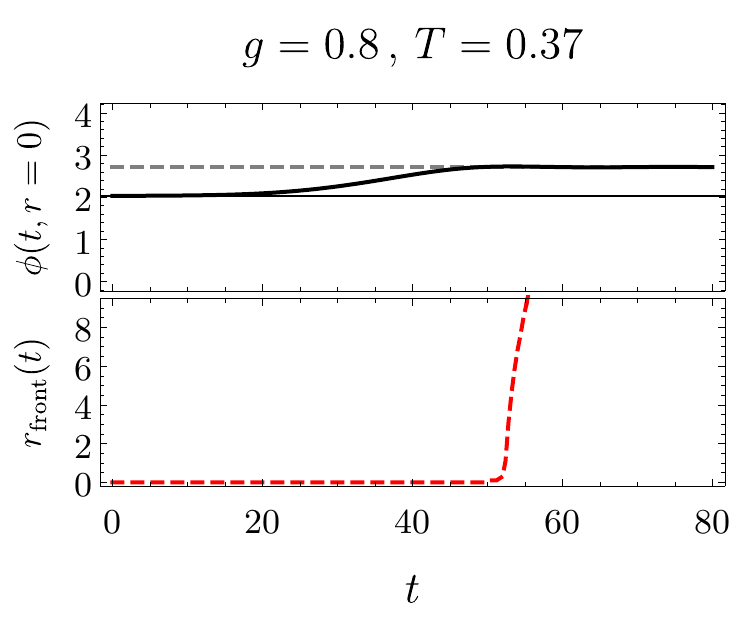}$\qquad\qquad$
\includegraphics[scale=0.9]{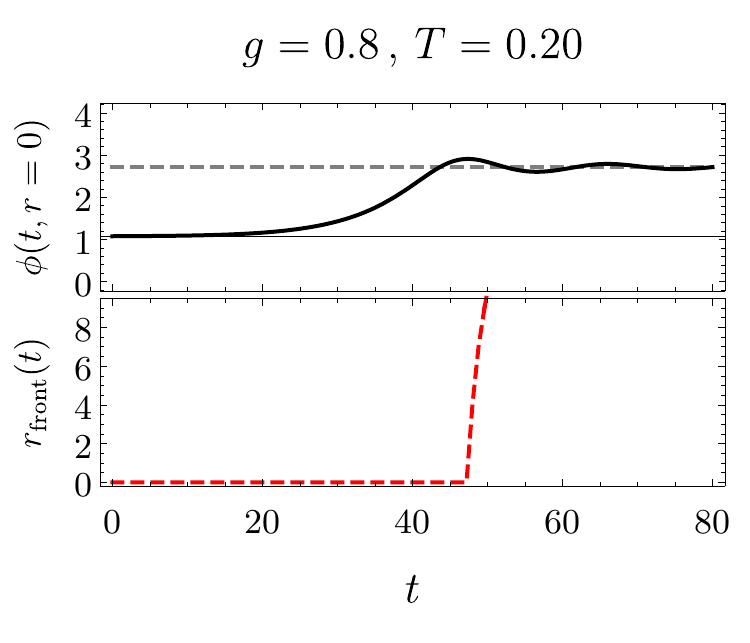}
\caption{{\it Time evolution of the wavefront point $r_\text{front}(t)$ (dotted red) as function of time. We define the $r_\text{front}(t)$ as the radius coordinate of the last (or only) peak of the bubble oscillations (see black dots in Fig.~(\ref{timeEvolutionPlot})) from left to right. The continuous black line is the value of the field at the origin $\phi(t,r=0)$ which oscillates around the true vacuum. The bubble starts expanding at the time value corresponding to the red cusp. }}
\label{frontWaveEvolution}
\end{center}
\end{figure}

The static solution can be treated as initial condition at $t=0$ for the time evolution described by Eq.~(\ref{timeDepSols}). We can consider a bubble moving uniformly after its nucleation by perturbing the static solution through a small time-independent velocity $\epsilon$ and solving Eq.~(\ref{timeDepSols}) with initial conditions
\begin{equation}
\phi(t=0,r) = \phi_{stat}(r)\,,\qquad \frac{\partial}{\partial t}\phi(t,r)\Big|_{t=0} = \epsilon\,,\qquad \frac{\partial}{\partial r}\phi(t,r)\Big|_{r=0} = 0.
\end{equation}
We plot in Fig.~(\ref{timeEvolutionPlot}) the time-evolution of the bounce solutions, that start oscillating around the true vacuum very quickly. As shown in Fig.~(\ref{frontWaveEvolution}), the wave starts propagating when $\phi(t,0)$ reaches the true vacuum value. At this time, corresponding to the red cups in Fig.~(\ref{frontWaveEvolution}), the radius $R(t)$ differs by the initial radius $R(t=0)$ by only $\mathcal{O}(1)$ coefficients and $R_0=R(t=0)$ can be used as a good estimate in the right-hand side of Eq.~(\ref{eq:criticalGamma}).
\vspace{3 cm}

\bibliographystyle{JHEP}
{\footnotesize
\bibliography{biblio}}

\providecommand{\href}[2]{#2}\begingroup\raggedright\begin{thebibliography}{10}

\bibitem{Abbott:2016blz}
{\bf LIGO Scientific, Virgo} Collaboration, B.~P. Abbott et~al. {\em Phys. Rev.
  Lett.} {\bf 116} (2016), no.~6 061102,
  [\href{http://arxiv.org/abs/1602.03837}{{\tt arXiv:1602.03837}}].

\bibitem{GBM:2017lvd}
{\bf LIGO Scientific, Virgo, Fermi GBM, INTEGRAL, IceCube, AstroSat Cadmium
  Zinc Telluride Imager Team, IPN, Insight-Hxmt, ANTARES, Swift, AGILE Team,
  1M2H Team, Dark Energy Camera GW-EM, DES, DLT40, GRAWITA, Fermi-LAT, ATCA,
  ASKAP, Las Cumbres Observatory Group, OzGrav, DWF (Deeper Wider Faster
  Program), AST3, CAASTRO, VINROUGE, MASTER, J-GEM, GROWTH, JAGWAR,
  CaltechNRAO, TTU-NRAO, NuSTAR, Pan-STARRS, MAXI Team, TZAC Consortium, KU,
  Nordic Optical Telescope, ePESSTO, GROND, Texas Tech University, SALT Group,
  TOROS, BOOTES, MWA, CALET, IKI-GW Follow-up, H.E.S.S., LOFAR, LWA, HAWC,
  Pierre Auger, ALMA, Euro VLBI Team, Pi of Sky, Chandra Team at McGill
  University, DFN, ATLAS Telescopes, High Time Resolution Universe Survey,
  RIMAS, RATIR, SKA South Africa/MeerKAT} Collaboration, B.~P. Abbott et~al.
  {\em Astrophys. J.} {\bf 848} (2017), no.~2 L12,
  [\href{http://arxiv.org/abs/1710.05833}{{\tt arXiv:1710.05833}}].

\bibitem{Bernard:2004je}
{\bf MILC} Collaboration, C.~Bernard, T.~Burch, E.~B. Gregory, D.~Toussaint,
  C.~E. DeTar, J.~Osborn, S.~Gottlieb, U.~M. Heller, and R.~Sugar {\em Phys.
  Rev.} {\bf D71} (2005) 034504,
  [\href{http://arxiv.org/abs/hep-lat/0405029}{{\tt hep-lat/0405029}}].

\bibitem{Aoki:2006we}
Y.~Aoki, G.~Endrodi, Z.~Fodor, S.~D. Katz, and K.~K. Szabo {\em Nature} {\bf
  443} (2006) 675--678, [\href{http://arxiv.org/abs/hep-lat/0611014}{{\tt
  hep-lat/0611014}}].

\bibitem{Cheng:2006qk}
M.~Cheng et~al. {\em Phys. Rev.} {\bf D74} (2006) 054507,
  [\href{http://arxiv.org/abs/hep-lat/0608013}{{\tt hep-lat/0608013}}].

\bibitem{Kajantie:1996mn}
K.~Kajantie, M.~Laine, K.~Rummukainen, and M.~E. Shaposhnikov {\em Phys. Rev.
  Lett.} {\bf 77} (1996) 2887--2890,
  [\href{http://arxiv.org/abs/hep-ph/9605288}{{\tt hep-ph/9605288}}].

\bibitem{Rummukainen:1998as}
K.~Rummukainen, M.~Tsypin, K.~Kajantie, M.~Laine, and M.~E. Shaposhnikov {\em
  Nucl. Phys.} {\bf B532} (1998) 283--314,
  [\href{http://arxiv.org/abs/hep-lat/9805013}{{\tt hep-lat/9805013}}].

\bibitem{Csikor:1998eu}
F.~Csikor, Z.~Fodor, and J.~Heitger {\em Phys. Rev. Lett.} {\bf 82} (1999)
  21--24, [\href{http://arxiv.org/abs/hep-ph/9809291}{{\tt hep-ph/9809291}}].

\bibitem{Sakharov:1967dj}
A.~D. Sakharov {\em Pisma Zh. Eksp. Teor. Fiz.} {\bf 5} (1967) 32--35. [Usp.
  Fiz. Nauk161,no.5,61(1991)].

\bibitem{Witten:1984rs}
E.~Witten {\em Phys. Rev.} {\bf D30} (1984) 272--285.

\bibitem{Cutting:2018tjt}
D.~Cutting, M.~Hindmarsh, and D.~J. Weir {\em Phys. Rev.} {\bf D97} (2018),
  no.~12 123513, [\href{http://arxiv.org/abs/1802.05712}{{\tt
  arXiv:1802.05712}}].

\bibitem{Kosowsky:1991ua}
A.~Kosowsky, M.~S. Turner, and R.~Watkins {\em Phys. Rev.} {\bf D45} (1992)
  4514--4535.

\bibitem{Kosowsky:1992vn}
A.~Kosowsky and M.~S. Turner {\em Phys. Rev.} {\bf D47} (1993) 4372--4391,
  [\href{http://arxiv.org/abs/astro-ph/9211004}{{\tt astro-ph/9211004}}].

\bibitem{Kosowsky:1992rz}
A.~Kosowsky, M.~S. Turner, and R.~Watkins {\em Phys. Rev. Lett.} {\bf 69}
  (1992) 2026--2029.

\bibitem{Jinno:2017fby}
R.~Jinno and M.~Takimoto {\em JCAP} {\bf 1901} (2019) 060,
  [\href{http://arxiv.org/abs/1707.03111}{{\tt arXiv:1707.03111}}].

\bibitem{Hindmarsh:2017gnf}
M.~Hindmarsh, S.~J. Huber, K.~Rummukainen, and D.~J. Weir {\em Phys. Rev.} {\bf
  D96} (2017), no.~10 103520, [\href{http://arxiv.org/abs/1704.05871}{{\tt
  arXiv:1704.05871}}].

\bibitem{Konstandin:2017sat}
T.~Konstandin {\em JCAP} {\bf 1803} (2018), no.~03 047,
  [\href{http://arxiv.org/abs/1712.06869}{{\tt arXiv:1712.06869}}].

\bibitem{Caprini:2015zlo}
C.~Caprini et~al. {\em JCAP} {\bf 1604} (2016), no.~04 001,
  [\href{http://arxiv.org/abs/1512.06239}{{\tt arXiv:1512.06239}}].

\bibitem{Kamionkowski:1993fg}
M.~Kamionkowski, A.~Kosowsky, and M.~S. Turner {\em Phys. Rev.} {\bf D49}
  (1994) 2837--2851, [\href{http://arxiv.org/abs/astro-ph/9310044}{{\tt
  astro-ph/9310044}}].

\bibitem{Kosowsky:2001xp}
A.~Kosowsky, A.~Mack, and T.~Kahniashvili {\em Phys. Rev.} {\bf D66} (2002)
  024030, [\href{http://arxiv.org/abs/astro-ph/0111483}{{\tt
  astro-ph/0111483}}].

\bibitem{Caprini:2009yp}
C.~Caprini, R.~Durrer, and G.~Servant {\em JCAP} {\bf 0912} (2009) 024,
  [\href{http://arxiv.org/abs/0909.0622}{{\tt arXiv:0909.0622}}].

\bibitem{Bodeker:2009qy}
D.~Bodeker and G.~D. Moore {\em JCAP} {\bf 0905} (2009) 009,
  [\href{http://arxiv.org/abs/0903.4099}{{\tt arXiv:0903.4099}}].

\bibitem{Bodeker:2017cim}
D.~Bodeker and G.~D. Moore {\em JCAP} {\bf 1705} (2017), no.~05 025,
  [\href{http://arxiv.org/abs/1703.08215}{{\tt arXiv:1703.08215}}].

\bibitem{Caprini:2019egz}
C.~Caprini et~al. {\em JCAP} {\bf 03} (2020), no.~03 024,
  [\href{http://arxiv.org/abs/1910.13125}{{\tt arXiv:1910.13125}}].

\bibitem{Croon:2018erz}
D.~Croon, V.~Sanz, and G.~White {\em JHEP} {\bf 08} (2018) 203,
  [\href{http://arxiv.org/abs/1806.02332}{{\tt arXiv:1806.02332}}].

\bibitem{Alanne:2019bsm}
T.~Alanne, T.~Hugle, M.~Platscher, and K.~Schmitz
  \href{http://arxiv.org/abs/1909.11356}{{\tt arXiv:1909.11356}}.

\bibitem{Breitbach:2018ddu}
M.~Breitbach, J.~Kopp, E.~Madge, T.~Opferkuch, and P.~Schwaller {\em JCAP} {\bf
  1907} (2019), no.~07 007, [\href{http://arxiv.org/abs/1811.11175}{{\tt
  arXiv:1811.11175}}].

\bibitem{Bian:2019szo}
L.~Bian, W.~Cheng, H.-K. Guo, and Y.~Zhang
  \href{http://arxiv.org/abs/1907.13589}{{\tt arXiv:1907.13589}}.

\bibitem{Hashino:2018zsi}
K.~Hashino, M.~Kakizaki, S.~Kanemura, P.~Ko, and T.~Matsui {\em JHEP} {\bf 06}
  (2018) 088, [\href{http://arxiv.org/abs/1802.02947}{{\tt arXiv:1802.02947}}].

\bibitem{Fairbairn:2019xog}
M.~Fairbairn, E.~Hardy, and A.~Wickens {\em JHEP} {\bf 07} (2019) 044,
  [\href{http://arxiv.org/abs/1901.11038}{{\tt arXiv:1901.11038}}].

\bibitem{Mohamadnejad:2019vzg}
A.~Mohamadnejad \href{http://arxiv.org/abs/1907.08899}{{\tt arXiv:1907.08899}}.

\bibitem{Dev:2019njv}
P.~S.~B. Dev, F.~Ferrer, Y.~Zhang, and Y.~Zhang
  \href{http://arxiv.org/abs/1905.00891}{{\tt arXiv:1905.00891}}.

\bibitem{Weinberg:1973am}
E.~J. Weinberg, {\em {Radiative corrections as the origin of spontaneous
  symmetry breaking}}.
\newblock PhD thesis, Harvard U., 1973.
\newblock \href{http://arxiv.org/abs/hep-th/0507214}{{\tt hep-th/0507214}}.

\bibitem{Curtin:2016urg}
D.~Curtin, P.~Meade, and H.~Ramani {\em Eur. Phys. J.} {\bf C78} (2018), no.~9
  787, [\href{http://arxiv.org/abs/1612.00466}{{\tt arXiv:1612.00466}}].

\bibitem{Weinberg:1974hy}
S.~Weinberg {\em Phys. Rev.} {\bf D9} (1974) 3357--3378.

\bibitem{Arnold:1992rz}
P.~B. Arnold and O.~Espinosa {\em Phys. Rev.} {\bf D47} (1993) 3546,
  [\href{http://arxiv.org/abs/hep-ph/9212235}{{\tt hep-ph/9212235}}]. [Erratum:
  Phys. Rev.D50,6662(1994)].

\bibitem{Coleman:1977py}
S.~R. Coleman {\em Phys. Rev.} {\bf D15} (1977) 2929--2936. [Erratum: Phys.
  Rev.D16,1248(1977)].

\bibitem{Linde:1980tt}
A.~D. Linde {\em Phys. Lett.} {\bf 100B} (1981) 37--40.

\bibitem{Linde:1981zj}
A.~D. Linde {\em Nucl. Phys.} {\bf B216} (1983) 421. [Erratum: Nucl.
  Phys.B223,544(1983)].

\bibitem{Wainwright:2011kj}
C.~L. Wainwright {\em Comput. Phys. Commun.} {\bf 183} (2012) 2006--2013,
  [\href{http://arxiv.org/abs/1109.4189}{{\tt arXiv:1109.4189}}].

\bibitem{Ellis:2018mja}
J.~Ellis, M.~Lewicki, and J.~M. No \href{http://arxiv.org/abs/1809.08242}{{\tt
  arXiv:1809.08242}}. [JCAP1904,003(2019)].

\bibitem{Espinosa:2010hh}
J.~R. Espinosa, T.~Konstandin, J.~M. No, and G.~Servant {\em JCAP} {\bf 1006}
  (2010) 028, [\href{http://arxiv.org/abs/1004.4187}{{\tt arXiv:1004.4187}}].

\bibitem{Guth:1979bh}
A.~H. Guth and S.~H.~H. Tye {\em Phys. Rev. Lett.} {\bf 44} (1980) 631.
  [Erratum: Phys. Rev. Lett.44,963(1980)].

\bibitem{Guth:1981uk}
A.~H. Guth and E.~J. Weinberg {\em Phys. Rev.} {\bf D23} (1981) 876.

\bibitem{Guth:1982pn}
A.~H. Guth and E.~J. Weinberg {\em Nucl. Phys. B} {\bf 212} (1983) 321--364.

\bibitem{Enqvist:1991xw}
K.~Enqvist, J.~Ignatius, K.~Kajantie, and K.~Rummukainen {\em Phys. Rev.} {\bf
  D45} (1992) 3415--3428.

\bibitem{Ellis:2019oqb}
J.~Ellis, M.~Lewicki, J.~M. No, and V.~Vaskonen {\em JCAP} {\bf 1906} (2019),
  no.~06 024, [\href{http://arxiv.org/abs/1903.09642}{{\tt arXiv:1903.09642}}].

\bibitem{Caprini:2009fx}
C.~Caprini, R.~Durrer, T.~Konstandin, and G.~Servant {\em Phys. Rev.} {\bf D79}
  (2009) 083519, [\href{http://arxiv.org/abs/0901.1661}{{\tt
  arXiv:0901.1661}}].

\bibitem{Moore:1995si}
G.~D. Moore and T.~Prokopec {\em Phys. Rev.} {\bf D52} (1995) 7182--7204,
  [\href{http://arxiv.org/abs/hep-ph/9506475}{{\tt hep-ph/9506475}}].

\bibitem{Moore:1995ua}
G.~D. Moore and T.~Prokopec {\em Phys. Rev. Lett.} {\bf 75} (1995) 777--780,
  [\href{http://arxiv.org/abs/hep-ph/9503296}{{\tt hep-ph/9503296}}].

\bibitem{Dorsch:2018pat}
G.~C. Dorsch, S.~J. Huber, and T.~Konstandin {\em JCAP} {\bf 1812} (2018),
  no.~12 034, [\href{http://arxiv.org/abs/1809.04907}{{\tt arXiv:1809.04907}}].

\bibitem{Konstandin:2014zta}
T.~Konstandin, G.~Nardini, and I.~Rues {\em JCAP} {\bf 1409} (2014), no.~09
  028, [\href{http://arxiv.org/abs/1407.3132}{{\tt arXiv:1407.3132}}].

\bibitem{Leitao:2014pda}
L.~Leitao and A.~Megevand {\em Nucl. Phys.} {\bf B891} (2015) 159--199,
  [\href{http://arxiv.org/abs/1410.3875}{{\tt arXiv:1410.3875}}].

\bibitem{Cornish:2001bb}
N.~J. Cornish {\em Phys. Rev.} {\bf D65} (2002) 022004,
  [\href{http://arxiv.org/abs/gr-qc/0106058}{{\tt gr-qc/0106058}}].

\bibitem{Moore:2014lga}
C.~J. Moore, R.~H. Cole, and C.~P.~L. Berry {\em Class. Quant. Grav.} {\bf 32}
  (2015), no.~1 015014, [\href{http://arxiv.org/abs/1408.0740}{{\tt
  arXiv:1408.0740}}].

\bibitem{PhysRevD.80.104009}
C.~Cutler and D.~E. Holz {\em Phys. Rev. D} {\bf 80} (Nov, 2009) 104009.

\bibitem{Aasi:2013wya}
{\bf KAGRA, LIGO Scientific, VIRGO} Collaboration, B.~P. Abbott et~al. {\em
  Living Rev. Rel.} {\bf 21} (2018), no.~1 3,
  [\href{http://arxiv.org/abs/1304.0670}{{\tt arXiv:1304.0670}}].

\bibitem{TheLIGOScientific:2014jea}
{\bf LIGO Scientific} Collaboration, J.~Aasi et~al. {\em Class. Quant. Grav.}
  {\bf 32} (2015) 074001, [\href{http://arxiv.org/abs/1411.4547}{{\tt
  arXiv:1411.4547}}].

\bibitem{Cornish:2018dyw}
T.~Robson, N.~J. Cornish, and C.~Liug {\em Class. Quant. Grav.} {\bf 36}
  (2019), no.~10 105011, [\href{http://arxiv.org/abs/1803.01944}{{\tt
  arXiv:1803.01944}}].

\bibitem{Graham:2017pmn}
{\bf MAGIS} Collaboration, P.~W. Graham, J.~M. Hogan, M.~A. Kasevich,
  S.~Rajendran, and R.~W. Romani \href{http://arxiv.org/abs/1711.02225}{{\tt
  arXiv:1711.02225}}.

\bibitem{Yagi:2011yu}
K.~Yagi, N.~Tanahashi, and T.~Tanaka {\em Phys. Rev.} {\bf D83} (2011) 084036,
  [\href{http://arxiv.org/abs/1101.4997}{{\tt arXiv:1101.4997}}].

\bibitem{Yagi:2013du}
K.~Yagi {\em Int. J. Mod. Phys.} {\bf D22} (2013) 1341013,
  [\href{http://arxiv.org/abs/1302.2388}{{\tt arXiv:1302.2388}}].

\bibitem{Sathyaprakash:2012jk}
B.~Sathyaprakash et~al. {\em Class. Quant. Grav.} {\bf 29} (2012) 124013,
  [\href{http://arxiv.org/abs/1206.0331}{{\tt arXiv:1206.0331}}]. [Erratum:
  Class. Quant. Grav.30,079501(2013)].

\end{thebibliography}\endgroup
\end{document}